\documentclass[10pt,pra,twocolumn,superscriptaddress,showpacs,floatfix]{revtex4-2}
\usepackage{graphics}
\usepackage{graphicx}
\usepackage{bm}
\usepackage{amsmath, upgreek}
\usepackage{amsfonts}
\usepackage{amssymb}
\usepackage{latexsym}
\usepackage[dvipsnames]{xcolor}
\usepackage{physics}
\usepackage{orcidlink}
\usepackage{hyperref}
\usepackage{braket}
\usepackage{soul}

\hypersetup{
    colorlinks=true,
    linkcolor=blue,
    filecolor=black,      
    urlcolor=blue,
    allcolors=blue,
}

\begin{document}

\title{Exact dynamics of emitter--mirror systems in one-dimensional waveguides}

\author{Mateusz Duda \orcidlink{0009-0008-3083-4686}}
\affiliation{School of Mathematical and Physical Sciences, University of Sheffield, Sheffield S3 7RH, United Kingdom}

\author{Thomas Hartwell \orcidlink{0009-0005-9799-9440}}
\affiliation{School of Chemical and Process Engineering, University of Leeds, Leeds LS2 9JT, United Kingdom}
\affiliation{School of Physics and Astronomy, University of Leeds, Leeds LS2 9JT, United Kingdom}

\author{Daniel Hodgson \orcidlink{0009-0005-9799-9440}}
\affiliation{School of Mathematical and Physical Sciences, University of Sheffield, Sheffield S3 7RH, United Kingdom}

\author{Gin Jose \orcidlink{0000-0001-9856-6755}}
\affiliation{School of Chemical and Process Engineering, University of Leeds, Leeds LS2 9JT, United Kingdom}

\author{Pieter Kok \orcidlink{0000-0002-6608-330X}}
\affiliation{School of Mathematical and Physical Sciences, University of Sheffield, Sheffield S3 7RH, United Kingdom}

\author{Almut Beige \orcidlink{0000-0001-7230-4220}}
\affiliation{School of Physics and Astronomy, University of Leeds, Leeds LS2 9JT, United Kingdom}


\begin{abstract}
Single-photon emitters in nanophotonic structures are a key building block for many photonic devices with quantum technology applications, including quantum sensors and quantum computers. In this paper, we therefore determine the exact dynamics of a two-level quantum emitter in a one-dimensional waveguide with a partially transparent mirror by directly solving the corresponding Schr\"odinger equation via a local-photon approach. In general, time-delay effects lead to non-Markovian dynamics and non-exponential decay profiles. The decay can resemble an exponential after a time that is much larger than the emitter--mirror round-trip time and becomes exponential in the Markovian limit, where the round-trip time is negligible. We also derive the spatial and spectral profiles of the emitted photon wave packet, showing how the spectrum is altered by the mirror.
\end{abstract}
\date{\today}

\maketitle


\section{Introduction}\label{sec:intro}

Quantum emitters are a crucial component of photonic quantum technologies, as they can be used to generate single photons on demand~\cite{OBrien2009, Wang2020, Kurtsiefer2000, Kuhn2002} (e.g., in distributed quantum computing~\cite{Senellart2026}). Furthermore, coupling quantum emitters to nanophotonic structures provides control over the photon emission process. This can be achieved, for example, via the use of a cavity that enhances the spontaneous emission rate~\cite{Liu2018, Phillips2024}, or a waveguide that confines photon propagation to one dimension and directs photons to a desired output port~\cite{Coles2016, Duda2026}. Single-photon emitters have been integrated with waveguide and cavity structures across a variety of emitter platforms, including semiconductor quantum dots~\cite{Lodahl2015}, superconducting qubits~\cite{Blais2021}, and color centers in diamond~\cite{Janitz2020}.

In many of these situations time-delay effects are negligible, and the dynamics of an emitter with two internal states can be described by a Markovian master equation of the general form~\cite{Manzano2020}
\begin{eqnarray} \label{eq:H2}
    \dot \rho_{\rm e}(t) &=& -i \left[ H_{\rm e},\rho_{\rm e}(t) \right] + \Gamma \, \sigma^- \rho_{\rm e}(t) \sigma^+ \notag \\
    && - \Gamma/2 \, \left\{ \sigma^+ \sigma^-, \rho_{\rm e}(t) \right\} ,
\end{eqnarray}
where $\rho_{\rm e}(t)$ and $\Gamma$ are the density matrix and the decay rate of the emitter, respectively, while $H_{\rm e}$ is the emitter Hamiltonian and $\sigma^\pm$ are the atomic raising and lowering operators. In the absence of laser driving, the emitter loses its excitation according to a simple exponential decay law~\cite{Eberly, Longhi, Hartwell2026, Wang2025}, with the corresponding emission rate $\Gamma$ depending on the nanophotonic structure surrounding the emitter~\cite{Liu2018, Phillips2024}.

\begin{figure}[t]
    \centering  \includegraphics[width=0.85\linewidth]{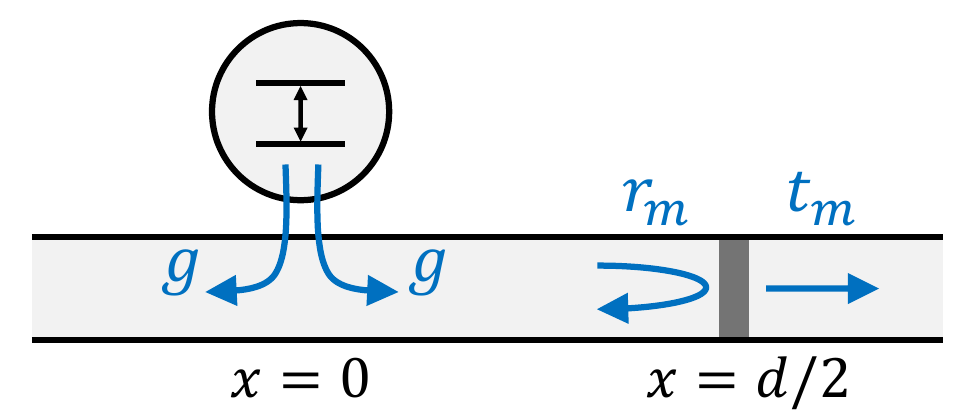}
    \caption{Diagram of a two-level quantum emitter coupled to a waveguide with a partially-transparent mirror with transmission and reflection coefficients $t_{\rm m}$ and $r_{\rm m}$, respectively. The emitter couples to the local electromagnetic field at ${x=0}$ with coupling rate $g$, and $d/2$ denotes the distance between the emitter and the mirror.}
    \label{fig:diagram}
\end{figure}

However, in the non-Markovian regime of waveguide quantum electrodynamics (QED), the emission process can become non-exponential~\cite{Pichler2016, Ramos2016, Barkemeyer2022, Cilluffo2026}. A paradigmatic example of such a system is a two-level quantum emitter coupled to a waveguide with a mirror, as illustrated in Fig.~\ref{fig:diagram}. If the photon round-trip time between the emitter and the mirror is comparable to the emission lifetime, time-delayed feedback produces non-trivial dynamics that can no longer be described by the Markovian master equation in Eq.~(\ref{eq:H2}). This is because the emitted light returns to its source before the emission process is completed, thereby interfering with the generation of the remaining photonic wave packet. Consequently, the dynamics of the emitter depends not only on its current state but also on its past evolution. It has been shown that Non-Markovian dynamics can be used for population trapping~\cite{Regidor2021}, generating emitter--emitter entanglement~\cite{Zhang2020} and photonic cluster states~\cite{Pichler2017}, exciting bound states~\cite{Calajo2019, Trivedi2021, Barkemeyer2021}, and signal amplification~\cite{Wiegand2021}. In addition, fluorescence lifetime measurements in non-Markovian systems can be used for photonic quantum sensing of the structured environment of an emitter, since the emission process depends on the properties of the surroundings. Notice that the emitter dynamics can be influenced by reflecting surfaces over distances that are not restricted by the wavelength of the emitted light; they are only limited by the length of the generated photonic wave packets.

The emitter--mirror system has been investigated theoretically and experimentally with superconducting circuits~\cite{Gu2017, Hoi2015}, ions~\cite{Drexhage1970, Amos1997, Eschner2001}, dye molecules~\cite{Frimmer2013}, and quantum dots~\cite{Qvotrup2025}, with interference-induced variations in decay rates being observed. In these experiments, the photon round-trip time is negligible and the dynamics of the emitter can be described by Eq.~(\ref{eq:H2}), with an emission rate $\Gamma$ that depends on the emitter--mirror distance due to interference effects~\cite{Barnes1998, Furtak-Wells2018, Ghamdi2026}. However, in general, time-delay effects that lead to non-Markovian dynamics need to be taken into account. Theoretically, such dynamics have been studied analytically using delay-differential equations~\cite{Cook1987, Dorner2002, Tufarelli2013, Tufarelli2014, Fang2018, Barkemeyer2022} and numerically using quantum-trajectory or matrix-product-state simulations~\cite{Crowder2020, Regidor2021, Crowder2022, Crowder2024, Pichler2016, Guimond2017}. 

While previous studies of emitter--mirror non-Markovian waveguide QED have considered a semi-infinite waveguide with a perfect mirror, this paper studies the more general case of mirrors with finite transmission and reflection coefficients, $t_{\rm m}$ and $r_{\rm m}$ (cf.~Fig.~\ref{fig:diagram}). We proceed similarly to Refs.~\cite{Eberly, Longhi, Hartwell2026, Wang2025}, which studied the emission from a single two-level system in free space by solving the Schr\"odinger equation of the emitter and field analytically. Here we solve the Schr\"odinger equation in position space using a locally-acting Hamiltonian $H$ of the form
\begin{eqnarray}\label{eq:H}
    H &=& H_{\rm e} + H_{\rm f} + H_{\text{int}},
\end{eqnarray}
where $H_{\rm e}$ and $H_{\rm f}$ denote the free Hamiltonian of the emitter and the quantized electromagnetic field, respectively, and $H_{\text{int}}$ is the emitter--field interaction Hamiltonian. As in Ref.~\cite{Hartwell2026}, we employ a Dyson series expansion that allow us to decompose the full evolution of the emitter and the field into local and time-delayed reabsorption processes, providing an intuitive picture for the dynamics. Moreover, by developing a quantum trajectory model for our system (Appendix~\ref{app:trajectories}), we show that our analytical results agree with well-established numerical simulation methods. The results are also consistent with the analytical results derived for the perfect mirror case using delay-differential equations~\cite{Dorner2002, Tufarelli2013, Tufarelli2014}.

The rest of this paper is organized as follows. In Section~\ref{sec:model}, we outline the theoretical tools that we use to model the dynamics of the system in Fig.~\ref{fig:diagram}. In Section~\ref{sec:calculations}, we calculate the full quantum state $\ket{\psi(t)}$ of the emitter and field via a Dyson series expansion. Then, in Section~\ref{sec:results}, we use our results to study observable properties like the population of the excited state of the emitter, as well as the photoluminescence profile and the spectrum of the emitted light. We conclude with a summary and discussion in Section~\ref{sec:conc}.


\section{Hamiltonian and Dyson series expansion} \label{sec:model}

In this section we introduce the methodology and notation that will be used throughout the paper. We start by writing the Hamiltonian $H$ in Eq.~(\ref{eq:H}) as ${H = H_0 + H_1}$ with 
\begin{equation}
    H_0 = H_{\rm e} + H_{\rm f}
\end{equation}
containing the free emitter and field terms (including the mirror Hamiltonian), while
\begin{equation}\label{eq:H_1}
    H_1 = H_{\text{int}}
\end{equation}
only contains the local emitter--field interaction. The above separation allows us to solve the resulting dynamics of the emitter--mirror system exactly by utilizing the Dyson series expansion~\cite{Waite2025,Hartwell2026}, which we present in Section~\ref{subsec:model_Dyson_series}. Afterwards, in Section~\ref{subsec:model_Hamiltonian}, we introduce the Hilbert space of local photon excitations and identify the free Hamiltonian $H_0$. Section~\ref{subsec:model_Hamiltonianb} introduces the interaction term $H_1$.


\subsection{Dyson series}\label{subsec:model_Dyson_series}

The time evolution operator associated with the time-independent Hamiltonian $H$ in Eq.~(\ref{eq:H}) is given by
\begin{equation} \label{U}
U(t,0) = e^{-iHt}
\end{equation}
(with ${\hbar = 1}$). The purpose of this subsection is to obtain an expression for $U(t,0)$ in terms of the operator $U_0(t,0)$ for the free evolution of the emitter--mirror system, which is given by 
\begin{equation}\label{E7}
U_0(t,0) = e^{-iH_0t},
\end{equation}
as well as the interaction Hamiltonian $H_1$. As we shall see in the following subsections, $U_0(t,0)$ evolves emitter--field states in a simple way, and we also know how to apply $H_1$ to these states. 

The Dyson series that we use in our calculations can be obtained by first moving into the interaction picture with respect to $H_0$, evaluating the time evolution operator in the interaction picture using the standard Dyson series expansion, and then returning to the Schr\"{o}dinger picture. As shown in Refs.~\cite{Waite2025, Hartwell2026}, $U(t,0)$ in Eq.~(\ref{U}) can be written as a sum of $U_n(t,0)$ terms: 
\begin{equation}\label{eq:Dyson_short}
    U(t,0) = \sum_{n=0}^{\infty} U_n(t,0).
\end{equation}
The first term in the series, $U_0(t,0)$, is the time evolution operator in Eq.~(\ref{E7}). In addition, the ${n=1}$ term is given by
\begin{equation}\label{eq:Dyson_short_1}
    U_1(t,0) = -i \int_0^t {\rm d}t_1\; U_0(t,t_1)\; H_1\; U_0(t_1,0),
\end{equation}
and the remaining terms with ${n\geq2}$ have the form
\begin{align}\label{eq:Dyson_short_n}
\begin{split}
    U_n(t,0) =&\; (-i)^n \int_0^t {\rm d}t_1 \ldots \int_0^{t_{n-1}} {\rm d}t_n\\
    & \times U_0(t,t_1)\; H_1\; U_0(t_1,t_2) \ldots H_1\; U_0(t_n,0),
\end{split}
\end{align}
with the time ordering ${t \geq t_1 \geq t_2 \geq \ldots \geq t_n \geq 0}$. Equation~(\ref{eq:Dyson_short_n}) shows that we can calculate the quantum state  
\begin{equation} \label{E9}
{\ket{\psi(t)} = U(t,0)\ket{\psi(0)}}
\end{equation}
of the emitter--field system at time $t$ by applying sequences of time evolution operators $U_0(t_i,t_{i+1})$ with $t_i \ge t_{i+1}$ and interaction Hamiltonians $H_1$ to the initial state $\ket{\psi(0)}$. To simplify the calculation of the relevant terms, let us point out that this can be done iteratively since the $U_n(t,0)$ terms obey the relation 
\begin{align}\label{eq:Dyson_short_n2}
\begin{split}
    U_{n+1} (t,0) =&\; -i \int_0^t {\rm d}t_1 \; U_0(t,t_1)\; H_1\; U_{n}(t_1,0).
\end{split}
\end{align}


\subsection{The free evolution $U_0(t_i,t_{i+1})$}\label{subsec:model_Hamiltonian}

In this paper, we consider a two-level quantum emitter coupled to a one-dimensional waveguide with a semi-transparent mirror at a distance $d/2$ away from the emitter, as shown in Fig.~\ref{fig:diagram}. The free Hamiltonian of the emitter can be written as
\begin{equation}
    H_{\rm e} = \omega_{\rm e} \sigma^+ \sigma^- ,
\end{equation}
where $\omega_{\rm e}$ is the transition frequency and  ${\sigma^+ = \ketbra{\rm e}{\rm g}}$ and ${\sigma^- = \ketbra{\rm g}{\rm e}}$ are the raising and lowering operators, respectively. Here $\ket{{\rm g}}$ is the ground state and $\ket{{\rm e}}$ is the excited state of the emitter. These states evolve under the time evolution operator $U_{\rm e}(t_i,t_{i+1})$ associated with $H_{\rm e}$ according to
\begin{align} \label{E11}
\begin{split}
    U_{\rm e}(t_i,t_{i+1})\ket{{\rm g}} =&\; \ket{{\rm g}},\\
    U_{\rm e}(t_i,t_{i+1})\ket{{\rm e}} =&\; e^{-i\omega_{\rm e}(t_i-t_{i+1})} \ket{{\rm e}}.
\end{split}
\end{align}
We now consider the Hamiltonian $H_{\rm f}$ of the quantized electromagnetic field in the waveguide. As was shown in Ref.~\cite{Waite2025}, this Hamiltonian is of the form
\begin{equation}\label{eq:H_f}
    H_{\rm f} =  H_{\rm dyn} +  H_{\rm m},
\end{equation}
with the first term describing free photon propagation and the second term accounting for reflections at the mirror. With photons propagating only in one dimension, the Hamiltonian $H_{\rm dyn}$ can be written as~\cite{Hodgson2022, Shen2005, Roy2017, Duda2026}
\begin{equation}\label{eq:H_dyn}
    H_{\rm dyn} = -ic \sum_{s=\pm1} s \int_{-\infty}^{\infty} {\rm d}x \; a_s^{\dag}(x) \frac{\partial}{\partial x} a_s(x),
\end{equation}
where $c$ is the speed of light in the waveguide, and $a_s^{\dag}(x)$ and $a_s(x)$ are the creation and annihilation operators for local electromagnetic field excitations at position $x$ moving in the $s$ direction, respectively, with ${s=+1}$ denoting the $+x$ direction (i.e., right-moving photons) and ${s=-1}$ denoting the $-x$ direction (i.e., left-moving photons).

The operators $a_s(x)$ and $a_s^{\dag}(x)$ allow us to represent all possible quantum states within the Hilbert space of the field. For example, when $a_s^{\dag}(x)$ acts on the vacuum state $\ket{0}$, we obtain the single-excitation state ${\ket{1_s(x)} = a_s^{\dag}(x)\ket{0}}$. This localized state is non-normalizable, so physical photon wave packets must be constructed by taking normalized superpositions of such states [i.e., by integrating over a suitable distribution $f(x)$]. For this reason, these states have previously been referred to as {\textit{blips}} (bosons localized in position)~\cite{Hodgson2022}. Using blips as building blocks for photon states in the position basis is analogous to constructing normalized wave packets in the frequency (or wave number) basis~\cite{Duda2024}. As blips located in different positions or moving in different directions are distinguishable, the blip operators obey the bosonic commutation relations
\begin{equation}
    \left[ a_s(x), a_{s'}^{\dagger}(x') \right] = \delta_{ss'} \delta(x-x').
\end{equation}

Using the above relation, one can show that the free Hamiltonian $H_{\rm dyn}$ in Eq.~(\ref{eq:H_dyn}) propagates photons at the speed of light, as we would expect from classical electromagnetism. In analogy to Eq.~(\ref{E11}), the corresponding time evolution operator $U_{\rm dyn}(t_i,t_{i+1})$ evolves zero- and one-excitation states of the field according to
\begin{align}\label{E14}
\begin{split}
    U_{\rm dyn}(t_i,t_{i+1}) \ket{0} =&\; \ket{0},\\
    U_{\rm dyn}(t_i,t_{i+1}) \ket{1_s(x)} =&\; \ket{1_s(x + sc(t_i-t_{i+1}))}.
\end{split}
\end{align}
We also need to consider the local mirror Hamiltonian $H_{\rm m}$ in Eq.~(\ref{eq:H_f}), which is given by~\cite{Waite2025}
\begin{equation}\label{eq:H_m}
    H_{\rm m} = J a_{-1}^{\dag}\left(d/2 \right) a_1\left( d/2 \right) + {\rm H.c.},
\end{equation}
where $J$ is the rate with which the mirror couples right- and left-moving photons and ${\rm H.c.}$ denotes the Hermitian conjugate. The above interaction can give rise to a change of the photon propagation direction when a photon reaches the mirror at ${x=d/2}$. As was shown in Ref.~\cite{Waite2025} using the Dyson series expansion from Section~\ref{subsec:model_Dyson_series}, the time evolution operator $U_{\rm f}(t_i,t_{i+1})$ associated with ${H_{\rm f} = H_{\rm dyn} + H_{\rm m}}$ evolves field states such that ${U_{\rm f}(t_i,t_{i+1})\ket{0} = \ket{0}}$ and
\begin{widetext}
\begin{align}\label{eq:one_photon_plus_x}
\begin{split}
    U_{\rm f}(t_i,t_{i+1})\ket{1_s(x)} =& \; 
    t_{\rm m} \; \Theta\left(s \left( d/2 -x \right) \right) \Theta\left( s \left( x - d/2 \right) + c(t_i-t_{i+1}) \right) \ket{1_s(x + sc(t_i-t_{i+1}))} \\
    & + r_{\rm m} \; \Theta\left(s \left( d/2 -x \right) \right) \Theta\left( s \left( x - d/2 \right) + c(t_i-t_{i+1}) \right) \ket{1_{-s}(d-x - sc(t_i-t_{i+1}))} \\
    & + \Theta \left(s\left(d/2 -x \right) \right) \Theta\left(s \left( d/2 - x \right) - c(t_i-t_{i+1})  \right) \ket{1_s(x + sc(t_i-t_{i+1}))} \\
    & + \Theta\left(s \left(x - d/2 \right) \right) \ket{1_s(x + sc(t_i-t_{i+1}))},
\end{split}
\end{align}
\end{widetext}
where $\Theta(x)$ is the Heaviside step function. Moreover, 
\begin{equation}\label{eq:t_r}
    t_{\rm m} = \frac{1 - \left( |J|/2c\right)^2}{1 + \left( |J|/2c\right)^2} \quad 
    \text{and} \quad r_{\rm m} = - \frac{iJ/c}{1 + \left( |J|/2c\right)^2}
\end{equation}
are the transmission and reflection coefficients of the mirror, respectively.

In our case, with the emitter located at ${x=0}$ and the mirror at ${x=d/2}$ with ${d>0}$ (see Fig.~\ref{fig:diagram}), we know that a photon emitted in the $-x$ direction will propagate freely without being affected by the mirror. This means that, at ${x=0}$, we have
\begin{subequations}
\begin{equation}\label{eq:one_photon_plus_x2}
    U_{\rm f}(t_i,t_{i+1})\ket{1_{-1}(0)} = \ket{1_{-1}(-c(t_i-t_{i+1}))}, 
\end{equation}
\begin{align}\label{eq:one_photon_plus_x3}
\begin{split}
    U_{\rm f} & (t_i,t_{i+1}) \ket{1_1(0)} \\
    =& \;  t_{\rm m} \; \Theta\left( c(t_i-t_{i+1}) - d/2 \right) \ket{1_{1}(c(t_i-t_{i+1}))} \\
    & \; + r_{\rm m} \; \Theta\left( c(t_i-t_{i+1}) - d/2 \right) \ket{1_{-1}(d - c(t_i-t_{i+1}))}  \\
    & \; + \Theta\left(d/2 - c(t_i-t_{i+1}) \right) \ket{1_1(c(t_i-t_{i+1}))}.
\end{split}
\end{align}
\end{subequations}
In Eq.~(\ref{eq:one_photon_plus_x3}), we see that a photon moving in the $+x$ direction propagates freely until it reaches the mirror, at which point it acquires transmitted and reflected components. Analogous equations can be derived for higher-excitation field states. However, in this paper we study single-photon emission and hence are only interested in the dynamics of single-excitation states $\ket{1_s(x)}$.

The Hamiltonians $H_{\rm e}$ and $H_{\rm f}$ of the emitter and field commute with each other, so the time evolution operator $U_0(t_i,t_{i+1})$ associated with ${H_0 = H_{\rm e} + H_{\rm f}}$, which appears in the Dyson series [Eq.~(\ref{eq:Dyson_short})], is simply the tensor product
\begin{equation}
    U_0(t_i,t_{i+1}) = U_{\rm e}(t_i,t_{i+1}) \otimes U_{\rm f}(t_i,t_{i+1}).
\end{equation}
Using Eqs.~(\ref{E11}), (\ref{eq:one_photon_plus_x2}), and (\ref{eq:one_photon_plus_x3}), we can therefore calculate how $U_0(t_i,t_{i+1})$ acts on both the emitter states $\{\ket{g},\ket{e}\}$ and the field vacuum and single-excitation states. In Section~\ref{sec:calculations}, we use these equations to obtain the dynamics of the emitter--mirror system in Fig.~\ref{fig:diagram}.


\subsection{The interaction Hamiltonian $H_1$}\label{subsec:model_Hamiltonianb}

To solve the Dyson series for the emitter--mirror system, we also need the Hamiltonian $H_1$ that describes the local emitter--field interaction at ${x=0}$~\cite{Hartwell2026}:
\begin{equation}\label{eq:H_int}
    H_1 = H_{\rm int} = \sum_{s=\pm1} g \; \sigma^- a_s^{\dag}(0) + {\rm H.c.}
\end{equation}
Here $g$ denotes the coupling rate between the emitter and the electromagnetic field in the waveguide (note that we assume equal coupling for ${s=\pm1}$). Now we can easily calculate how $H_1$ acts on the emitter states $\{\ket{{\rm g}}, \ket{{\rm e}}\}$ and the field states. Using Eq.~(\ref{eq:H_int}), we find that
\begin{align}\label{E20}
\begin{split}
H_1 \ket{{\rm e}} \ket{\psi_{\rm f}} =& \sum_{s=\pm1} g \ket{{\rm g}} a_s^\dag(0) \ket{\psi_{\rm f}},\\
H_1 \ket{{\rm g}} \ket{\psi_{\rm f}} =& \sum_{s=\pm1} g^* \ket{{\rm e}} a_s(0) \ket{\psi_{\rm f}},
\end{split}
\end{align}
given a general field state $\ket{\psi_{\rm f}}$. Hence, applying $H_1$ to an emitter--field state yields zero unless the emitter is in its excited state or a field excitation is present at ${x=0}$.


\section{Coherent dynamics of emitter and field} \label{sec:calculations}

For the purposes of modeling single-photon emission, we consider an initial state $ \ket{\psi(0)}$ where the emitter is excited and the field is in the vacuum state:
\begin{equation}\label{eq:initial_state}
    \ket{\psi(0)} = \ket{{\rm e}} \otimes \ket{0} = \ket{{\rm e},0}.
\end{equation}
The aim of this section is to find the full time evolution of the system as the emitter transitions into its ground state by emitting a photon. In order to calculate $\ket{\psi(t)}$ in Eq.~(\ref{E9}), we need to find how each $U_n(t,0)$ operator in Eq.~(\ref{eq:Dyson_short}) acts on $\ket{\psi(0)}$. As we shall see below, for all even $n$, the state $U_n(t,0) \ket{{\rm e},0}$ coincides with $\ket{{\rm e},0}$ up to an overall factor, since the interaction Hamiltonian $H_1$ is applied an even number of times. However, when $n$ is odd, $U_n(t,0) \ket{{\rm e},0}$ is a state with the emitter in the ground state $\ket{{\rm g}}$ and the waveguide containing a photon. In Section~\ref{subsec:calculations_1}, we show explicitly how the first three terms of the Dyson series are derived. Afterwards, we write the state vector $\ket{\psi(t)}$ of the system at time $t$ as a sum of even and odd terms,
\begin{equation}\label{eq:final_state}
    \ket{\psi(t)} = \ket{\psi_{\rm even}(t)} + \ket{\psi_{\rm odd}(t)},
\end{equation}
which we calculate separately using an iterative approach based on Eq.~(\ref{eq:Dyson_short_n2}). The even term $\ket{\psi_{\rm even}(t)}$ is derived in Section~\ref{subsec:calculations_2}, while the odd term $\ket{\psi_{\rm odd}(t)}$ is derived in Section~\ref{subsec:calculations_3}.


\subsection{First three terms of the Dyson series}\label{subsec:calculations_1}

\begin{figure}[b]
    \centering
    \includegraphics[width=\linewidth]{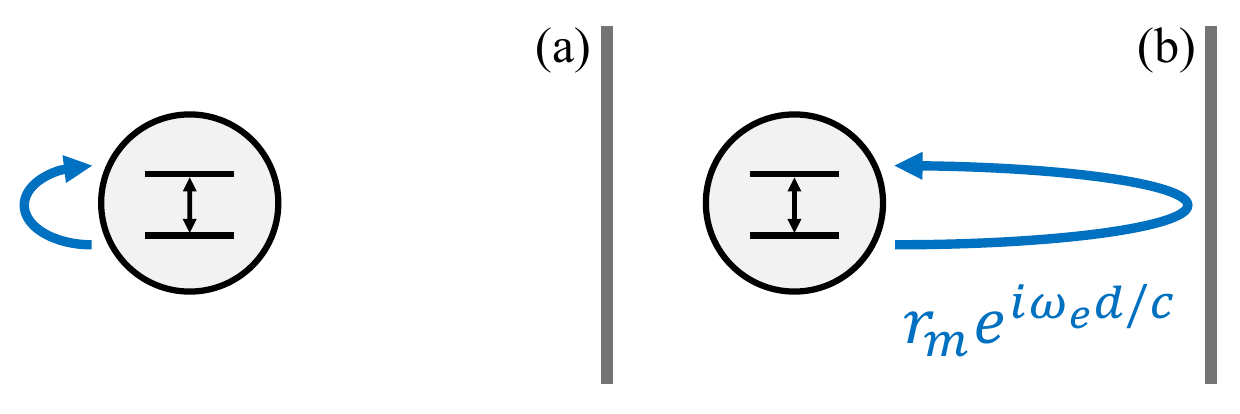}
    \caption{Reabsorption processes contributing to the Dyson series term $U_2(t,0)\ket{{\rm e},0}$ in Eq.~(\ref{eq:U_2}). (a) Instant reabsorption at ${x=0}$, corresponding to the first term in Eq.~(\ref{eq:U_2}). (b) Reabsorption after the round-trip time $d/c$, corresponding to the second term in Eq.~(\ref{eq:U_2}). In this case the amplitude of the state acquires the mirror reflection coefficient $r_{\rm m}$ and the round-trip phase $e^{i\omega_{\rm e} d/c}$.}
    \label{fig:one_loop}
\end{figure}

From Eq.~(\ref{E11}), it follows that the first term of the Dyson series, $U_n(t,0)\ket{{\rm e},0}$ with ${n=0}$, is given by
\begin{equation}\label{eq:U_0}
    U_0(t,0) \ket{{\rm e},0} = e^{-i\omega_{\rm e} t} \ket{{\rm e},0}.
\end{equation}
As one can see from Eq.~(\ref{eq:Dyson_short_n2}), to obtain the second term, we have to apply the interaction Hamiltonian $H_1$ once to the state in Eq.~(\ref{eq:U_0}), causing the emitter to de-excite while creating a local field excitation at ${x=0}$. This excitation then evolves under $U_0(t,t_1)$, and the ${s=1}$ contribution eventually acquires transmitted and reflected components. After using Eqs.~(\ref{eq:one_photon_plus_x2}), (\ref{eq:one_photon_plus_x3}), and (\ref{E20}), we find that
\begin{widetext} 
\begin{align} \label{eq:U_1}
\begin{split}
    U_1(t,0) \ket{{\rm e},0} 
    =& -ig \int_0^t {\rm d} t_1\; e^{-i\omega_{\rm e} t_1} \left[ \ket{{\rm g}, 1_{-1}(- c(t-t_1))} 
    +  t_{\rm m} \; \Theta\left( c(t-t_1) - d/2 \right) \ket{{\rm g},1_{1}(c(t-t_1))} \right.  \\
    &\hspace{0.7in} \left. + r_{\rm m} \; \Theta\left( c(t-t_1) - d/2 \right) \ket{{\rm g},1_{-1}(d - c(t-t_1))} 
    + \Theta\left(d/2 - c(t-t_1)  \right) \ket{{\rm g},1_1(c(t-t_1))} \right] .
\end{split}
\end{align}
\end{widetext}
Next we obtain the third term, $U_2(t,0) \ket{{\rm e},0}$, by using Eq.~(\ref{eq:Dyson_short_n2}) again and noticing that the interaction Hamiltonian can only excite the emitter by absorbing a photon at ${x=0}$. The term in Eq.~(\ref{eq:U_1}) containing the transmission coefficient $t_{\rm m}$ does not contribute to reabsorption since a photon that has been transmitted through the mirror never returns to the emitter. Calculating $U_2(t,0) \ket{{\rm e},0}$ using Eq.~(\ref{eq:Dyson_short_n2}) therefore gives
\begin{align}\label{eq:U_2}
\begin{split}
    U_2(t,0) \ket{{\rm e},0}
    =& - \frac{|g|^2}{c} \;  e^{-i \omega_{\rm e} t}  \; \ket{{\rm e},0} \int_0^t {\rm d} t_1 \\
    & \times  \left[ 1 + r_{\rm m} e^{i\omega_{\rm e} d/c} \; \Theta \left( t_1- d/c \right)  \right] .
\end{split}
\end{align}
The first term in Eq.~(\ref{eq:U_2}) corresponds to an instant reabsorption process. The second term corresponds to a delayed reabsorption after one round trip between the emitter and the mirror, and contains the round-trip time $d/c$ and the reflection coefficient $r_{\rm m}$. These processes are illustrated schematically in Fig.~\ref{fig:one_loop}. Performing the time integration leads to
\begin{align}\label{eq:U_2more}
\begin{split}
    U_2(t,0) \ket{{\rm e},0}
    =& - \frac{|g|^2}{c} \; e^{-i \omega_{\rm e} t} \; \ket{{\rm e},0} \\
    & \times \left[ t + r_{\rm m} e^{i\omega_{\rm e} d/c} \; \Theta \left( t- d/c \right) \left( t- d/c \right) \right] .
\end{split}
\end{align}
The step function in Eq.~(\ref{eq:U_2more}) indicates that the second term is only present if ${t>d/c}$, i.e., when the photon has had time to return to the emitter after being reflected. In the following two subsections, we calculate the higher order terms in the Dyson series, $U_n(t,0)\ket{{\rm e},0}$ for ${n>2}$, using an iterative approach and eventually obtain an exact expression for the full state ${\ket{\psi(t)} = U(t,0)\ket{{\rm e}, 0}}$.


\subsection{Higher order terms with even $n$} \label{subsec:calculations_2}

The calculations in the previous subsection show that $U_0(t,0)\ket{{\rm e},0}$ and $U_2(t,0)\ket{{\rm e},0}$ are both multiples of the initial state $\ket{{\rm e},0}$. The same is true for all higher order terms $U_n(t,0)\ket{{\rm e},0}$ with $n$ even, since applying an even number of interaction Hamiltonians $H_1$ returns the emitter to its excited state. This means that we can write the state vectors $U_n(t,0)\ket{{\rm e},0}$ for all even $n$ as
\begin{equation} \label{E30}
U_n(t,0)\ket{{\rm e},0} = c_n(t) \; e^{-i\omega_{\rm e} t} \ket{{\rm e},0},
\end{equation}
where $c_n(t)$ is a complex coefficient. When we substitute Eq.~(\ref{E30}) into Eq.~(\ref{eq:Dyson_short_n2}) to calculate $U_{n+1}(t,0)\ket{{\rm e},0}$, we find
\begin{widetext} 
\begin{align} \label{eq:U_1extra}
\begin{split}
    U_{n+1}(t,0) \ket{{\rm e},0} 
    =& -ig \int_0^t {\rm d} t_1\; c_n(t_1) \; e^{-i\omega_{\rm e} t_1} \left[ \ket{{\rm g}, 1_{-1}(- c(t-t_1))} 
    + t_{\rm m} \; \Theta\left( c(t-t_1) - d/2 \right) \ket{{\rm g},1_{1}(c(t-t_1))} \right. \\
    &\hspace{0.7in} \left.+ r_{\rm m} \; \Theta\left( c(t-t_1) - d/2 \right) \ket{{\rm g},1_{-1}(d - c(t-t_1))} 
    + \Theta\left(d/2 - c(t-t_1)  \right) \ket{{\rm g},1_1(c(t-t_1))}  \right] .
\end{split}
\end{align}
Like the state in  Eq.~(\ref{eq:U_1}), this equation contains terms describing a field excitation traveling to the left and to the right, as well as transmitted and reflected components that are present when the photon has had enough time to reach the mirror. Substituting Eq.~(\ref{eq:U_1extra}) into Eq.~(\ref{eq:Dyson_short_n2}) to calculate $U_{n+2}(t,0)\ket{{\rm e},0}$ leads to
\begin{align}\label{eq:U_2extra}
\begin{split}
    U_{n+2}(t,0) \ket{{\rm e},0}
    =& - \frac{|g|^2}{c} e^{-i \omega_{\rm e} t} \int_0^t {\rm d} t_1 \left[ c_n(t_1) + r_{\rm m} e^{i\omega_{\rm e} d/c} \; c_n(t_1-d/c) \; \Theta \left( t_1- d/c \right) \right] \ket{{\rm e},0},
\end{split}
\end{align}
\end{widetext}
in analogy with Eq.~(\ref{eq:U_2}). The only difference is that the integrand now also depends on the coefficients $c_n(t_1)$ and $ c_n(t_1-d/c)$. 

The above calculations show that the states $U_{n+2}(t,0) \ket{{\rm e},0}$ and $U_n(t,0) \ket{{\rm e},0}$ only differ by an additional emission and absorption process. One of these processes happens immediately, while the other involves a round trip between the emitter and the mirror, leading to the factor $r_{\rm m}e^{i\omega_e d/c}$ (see Fig.~\ref{fig:one_loop}). Since ${U_{n+2}(t,0)\ket{{\rm e},0} = c_{n+2}(t)e^{-i\omega_e t}\ket{{\rm e},0}}$ [see Eq.~(\ref{E30})], from Eq.~(\ref{eq:U_2extra}) it follows that the coefficient $c_{n+2}(t)$ is related to $c_n(t)$ via
\begin{align}\label{eq:c_n+2}
\begin{split}
    c_{n+2}(t) 
    =& - \frac{|g|^2}{c}  \left[ \int_0^t {\rm d} t_1 \; c_n(t_1) + r_{\rm m}  e^{i\omega_{\rm e} d/c} \Theta \left( t- d/c\right) \right. \\
    &\hspace{1.3in} \left. \times  \int_0^{t-d/c} {\rm d} t_1 \; c_n(t_1) \right],
\end{split}
\end{align}
after simplifying the second integral. This iterative equation can be used to calculate all contributions $U_n(t,0)\ket{{\rm e},0}$ when $n$ is even. From Eq.~(\ref{eq:U_0}), we can see that $c_0(t)=1$ for all times $t$. Consequently, 
\begin{align}\label{eq:U_2moremore}
\begin{split}
    c_2(t) 
    =& - \frac{|g|^2}{c}  \left[ t + r_{\rm m} e^{i\omega_{\rm e} d/c} \; \Theta \left( t- d/c \right) \left( t-d/c \right) \right],
\end{split}
\end{align}
in agreement with Eq.~(\ref{eq:U_2more}). Continuing this iteration and calculating the coefficients $c_n(t)$ with ${n>2}$ as described in Appendix~\ref{AppA} eventually leads us to the general expression 
\begin{align} \label{eq:binomial_series}
\begin{split}
    c_n(t) =& \; (-1)^{n/2} \; \frac{|g|^{n}}{(n/2)! \; c^{n/2}} \; \sum_{k=0}^{n/2} \binom{n/2}{k} \\
    & \times \left( r_{\rm m} \; e^{i\omega_{\rm e} d/c} \right)^k \Theta \left( t-kd/c \right) \left( t- kd/c \right)^{n/2},
\end{split}
\end{align}
for even $n$ and with ${\binom{n}{k} = n!/k!(n-k)!}$.

Let us now summarize the contribution $\ket{\psi_{\rm even}(t)}$ to the full state vector $\ket{\psi(t)}$, which contains the terms in which the emitter is in its excited state at time $t$, while the waveguide is in the vacuum state. To calculate $\ket{\psi_{\rm even}(t)}$, we need to sum over all even $n$. Introducing a new variable ${m=n/2}$ then leads us to
\begin{align} \label{eq:binomial_series2}
\begin{split}
    \ket{\psi_{\rm even}(t)} =& \sum_{n\hspace{1.5pt}\rm{even}} U_n(t,0)\ket{{\rm e}, 0} \\
    =&\; e^{-i \omega_{\rm e} t} \ket{{\rm e},0} \sum_{m=0}^{\infty} \sum_{k=0}^{m}  
    \frac{1}{k!(m-k)!} \left( - \frac{|g|^2}{c} \right)^m \\
    & \times \left( r_{\rm m} \; e^{i\omega_{\rm e} d/c} \right)^k \Theta \left( t-kd/c \right) \left( t- kd/c \right)^m.
\end{split}
\end{align}
Although this is the full result, we can express it in a more intuitive form by changing the order of the summations. In particular, we can rearrange the terms in Eq.~(\ref{eq:binomial_series2}) such that $m$ runs from $k$ to infinity and the sum over $k$ starts at ${k=0}$ and runs to infinity. Then, after substituting ${j = m-k}$, Eq.~(\ref{eq:binomial_series2}) becomes
\begin{align} \label{eq:binomial_series2b}
\begin{split}
     \ket{\psi_{\rm even}(t)} =& \; e^{-i \omega_{\rm e} t} \ket{{\rm e},0} \sum_{k=0}^\infty \sum_{j=0}^{\infty}  
    \frac{1}{j!k!} \left( - \frac{|g|^2}{c} \right)^{j+k}  \\
    & \times \left( r_{\rm m} \; e^{i\omega_{\rm e} d/c} \right)^k \Theta \left( t-kd/c \right) \left( t- kd/c \right)^{j+k}.
\end{split}
\end{align}
The sum over $j$ gives us the familiar expansion for the exponential function. Hence,
\begin{align} \label{eq:binomial_series4}
\begin{split}
\ket{\psi_{\rm even}(t)} =& \; e^{-i \Omega t} \sum_{k=0}^{\infty} \frac{a^k}{k!} \; \Theta \left( t-kd/c \right) \left( t- kd/c \right)^{k} \ket{{\rm e},0},
\end{split}
\end{align}
where
\begin{align} \label{condiextra2}
\begin{split}
& a = - r_{\rm m} \;  e^{i \Omega d/c} \; \Gamma /2 ,\\
& \Omega = \omega_{\rm e} - i \Gamma/2,~~ \Gamma = 2|g|^2/c .
\end{split}
\end{align}
Here $\Gamma$ denotes the free-space decay rate of the emitter and $\Omega$ is a complex frequency~\cite{Hartwell2026}. Equation~(\ref{eq:binomial_series4}) is an exact analytical result for the dynamics of the emitter, and is consistent with the semi-infinite waveguide results from Refs.~\cite{Dorner2002, Tufarelli2013, Tufarelli2014} obtained using delay-differential equations. The result nicely takes into account the physical processes that contribute to the emission of a photon in the presence of the mirror. However, it is also relatively complex due to the step functions that arise because there is a finite number of round trips that the photon can complete between the emitter and the mirror up to a given time $t$.

In addition to the intuitive result given in Eq.~(\ref{eq:binomial_series4}), a more succinct representation of the solution can be found. In particular, we show in Appendix~\ref{app:lambert_W} that the state $\ket{\psi_{\rm even}(t)}$ can also be written in the form
\begin{equation} \label{W1}
\ket{\psi_{\rm even}(t)} = e^{-i\Omega t}\sum_{k=-\infty}^{\infty} \frac{e^{[W_k(ad/c)]ct/d}}{1+W_k(ad/c)}\,\ket{{\rm e},0},
\end{equation}
where $W_k$ is a branch of the Lambert $W$ function~\cite{Sinha2020}. The Lambert $W$ function is multi-valued and analytic on the entire complex plane. There is a countably infinite set of branches, labeled by the integer $k$, on which the Lambert $W$ function is single-valued. Next we shall show that in the long-time limit where $t \gg d/c$, only the principal branch $W_0$ contributes to the solution.

To consider long times, we return to Eq.~(\ref{eq:binomial_series4}) and assume that
\begin{equation} \label{condi}
t \gg d/c \,,
\end{equation}
setting ${\Theta(t-kd/c) = 1}$ for all $k$. This assumption is well justified when $d/c$ is small enough that the terms in Eq.~(\ref{eq:binomial_series4}) where ${t-kd/c}$ becomes negative have a negligible contribution due to the $1/k!$ scaling. Proceeding as shown in Appendix~\ref{AppB}, we can now simplify the expression in Eq.~(\ref{eq:binomial_series4}) by summing over all $k$, which leads to 
\begin{align} \label{eq:exp_series3}
\begin{split}
\ket{\psi_{\rm even}(t)} =& \; \xi_0 \; e^{-i \Omega t} \; e^{\xi t} \; \ket{{\rm e},0}\\
=& \; \xi_0 \; e^{-i \omega_e t} e^{-(\Gamma - 2\xi)t/2} \; \ket{{\rm e},0}.
\end{split}
\end{align}
This equation shows that the emitter decays exponentially when ${t \gg d/c}$ (provided that solutions exist for $\xi_0$ and $\xi$), at a rate modified from the free-space decay rate $\Gamma$ by a constant $\xi$. Here
\begin{align} \label{condiextra3}
\begin{split}
\xi_0 =& \; \sum_{k=0}^\infty \frac{(-k)^k}{k!} \; (ad/c)^k
\end{split}
\end{align}
is the ${t \rightarrow 0}$ limit of the long-time solution in Eq.~(\ref{eq:exp_series3}), and $\xi$ is given by the solution to the relation
\begin{align} \label{condiextra66}
\begin{split}
\xi \; e^{\xi d/c} =& \; a.
\end{split}
\end{align}
For example, in the limit ${d \rightarrow 0}$, we have ${\xi_0 = 1}$ and ${\xi = a = - r_{\rm m} \Gamma/2}$.

We can obtain a result for $\xi_0$ by performing the summation in Eq.~(\ref{condiextra3}). To do this, we use the equality
\begin{equation}
\label{sumidentity}
\sum_{k=1}^{\infty}\frac{k^{k-1}}{k!}\left(-\frac{adx}{c}\right)^k = -W_0(adx/c).
\end{equation}
Our solution is entirely within the principal branch as the infinite sum only converges when ${|ad/c| = |r_{\rm m}|(\Gamma d/2c){\rm exp}(\Gamma d/2c) < 1/e}$. Hence, in the presence of a highly reflective mirror, an exponential solution only exists when the separation between the emitter and the mirror is very small or when the decay time of the emitter is very long, as expected in this approximation. The equality (\ref{sumidentity}) can be determined using the generating function for the Abel polynomials. If we take the $x$ derivative of both sides of the equality and then set $x=1$, we find that
\begin{equation}
\label{sumidentity2}
\sum_{k=1}^{\infty}\frac{k^k}{k!}\;\left(-ad/c\right)^k = -\frac{W_0(ad/c)}{1+W_0(ad/c)}\, ,
\end{equation}
which is identical to Eq.~(\ref{condiextra3}) up to the $k=0$ term. Noting that the $k=0$ term in the series is one, we can then show that
\begin{equation}
\label{finalxi0}
\xi_0 = \frac{1}{1+W_0(ad/c)}\, .
\end{equation}
Finally, it can be seen from Eq.~(\ref{condiextra66}) that $\xi = (c/d){\rm }W_0(ad/c)$, leading to an expression for $\ket{\psi_{\rm even}(t)}$ equal to the $W_0$ contribution of the full solution (\ref{W1}).

We discuss examples where solutions for $\xi$ and $\xi_0$ can be found in Section~\ref{subsec:results_large_times}. In addition, in Section~\ref{subsec:results_approximate} we show that completely neglecting the finite time delay $d/c$ in Eq.~(\ref{eq:binomial_series4}) leads to an exponential decay for all times $t$, which coincides with the Markovian approximation.


\subsection{Higher order terms with odd $n$}\label{subsec:calculations_3}

To obtain the $U_n(t,0)\ket{{\rm e},0}$ terms for odd $n$, we now return to Eqs.~(\ref{eq:Dyson_short_n2}) and (\ref{eq:final_state}) and notice that
\begin{align}\label{eq:Dyson_short_n99}
\begin{split}
    \ket{\psi_{\rm odd}(t)} =&\; -i \int_0^t {\rm d}t_1 \; U_0(t,t_1)\; H_1\; \ket{\psi_{\rm even}(t_1)} .
\end{split}
\end{align}
Substituting the exact solution for $\ket{\psi_{\rm even}(t)}$ in Eq.~(\ref{eq:binomial_series4}) into this equation, and again using Eqs.~(\ref{eq:one_photon_plus_x2}), (\ref{eq:one_photon_plus_x3}) and (\ref{E20}), we find that 
\begin{align} \label{eq:binomial_series44}
\begin{split}
\ket{\psi_{\rm odd}(t)} =& \; -ig \int_0^t {\rm d}t_1 \;  e^{-i \Omega t_1} \sum_{k=0}^{\infty} \frac{a^k}{k!} \; \Theta \left( t_1 - kd/c \right)  \\
     & \hspace*{-0.5cm} \; \times \left[  t_{\rm m} \; \Theta\left( c(t-t_1) - d/2 \right) \ket{{\rm g},1_{1}(c(t-t_1))} \right. \\
     & \hspace*{-0.5cm} \; + r_{\rm m} \; \Theta\left( c(t-t_1) - d/2 \right) \ket{{\rm g},1_{-1}(d - c(t-t_1))} \\
     & \hspace*{-0.5cm} \; + \Theta\left(d/2 - c(t-t_1) \right) \ket{{\rm g},1_1(c(t-t_1))} \\
     & \hspace*{-0.5cm} \; \left. + \ket{{\rm g},1_{-1}(-c(t-t_1))} \right]  \left( t_1- kd/c \right)^{k}.
\end{split}
\end{align}
Here the emitter is always in its ground state $\ket{{\rm g}}$, and the waveguide contains a single photon. There are two terms corresponding to an emitted photon moving away from the emitter (left or right), and a further two terms where the right-moving component splits into transmitted and reflected contributions if the time $t$ is sufficiently large to allow the photon to reach the mirror. Together, Eqs.~(\ref{eq:binomial_series4}) and (\ref{eq:binomial_series44}) provide the full solution for the state ${\ket{\psi(t)} = U(t,0)\ket{{\rm e},0}}$ of the system in Fig.~\ref{fig:diagram} [via Eq.~(\ref{eq:final_state})].

Suppose we again consider large times $t$ such that Eq.~(\ref{condi}) holds. Then the same is true for almost all times $t_1 \in (0,t)$ and we can replace the step functions in Eq.~(\ref{eq:binomial_series44}) either by zero or by one to a good approximation. Doing so, the state $\ket{\psi_{\rm odd}(t)}$ simplifies to 
\begin{align} \label{eq:binomial_series45}
\begin{split}
\ket{\psi_{\rm odd}(t)} =& \; -ig \int_0^t {\rm d}t_1 \;  e^{-i \Omega t_1} \; \sum_{k=0}^{\infty} \frac{a^k}{k!} \left( t_1- kd/c \right)^{k}\\
     & \hspace{0.8cm} \times \left[ t_{\rm m} \ket{{\rm g},1_{1}(c(t-t_1))}  \right.\\
     & \left. \hspace{1.25cm} +\; r_{\rm m} \ket{{\rm g},1_{-1}(d - c(t-t_1))} \right.\\
     & \left. \hspace{1.25cm} + \ket{{\rm g},1_{-1}(-c(t-t_1))} \right].
\end{split}
\end{align}
Using the results of Appendix~\ref{AppB}, we can express this state in terms of the constants $\xi_0$ and $\xi$ as
\begin{align}\label{eq:one_photon_plus_x5}
\begin{split}
     \ket{\psi_{\rm odd}(t)} =& \; - ig \xi_0 
     \int_0^t {\rm d}t_1 \; e^{-i \Omega t_1} \; e^{\xi t_1} \left[ t_{\rm m} \ket{{\rm g},1_{1}(c(t-t_1))}  \right.\\
     & \left. \hspace{1.5cm} +\; r_{\rm m} \ket{{\rm g},1_{-1}(d - c(t-t_1))} \right.\\
     & \left. \hspace{1.5cm} + \ket{{\rm g},1_{-1}(-c(t-t_1))} \right].
\end{split}
\end{align}
Notice that the same state is obtained after substituting $\ket{\psi_{\rm even}(t)}$ in Eq.~(\ref{eq:exp_series3}) into Eq.~(\ref{eq:Dyson_short_n99}).


\section{Observable consequences and discussion}\label{sec:results}

In this section, we use the analytical results of Section~\ref{sec:calculations} to study the dynamics of the two-level emitter and the emitted photon. First, in Section~\ref{subsec:results_exact}, we visualize the exact emitter dynamics based on Eq.~(\ref{eq:binomial_series4}), demonstrating how the mirror affects the emission process in different parameter regimes. Then, in Section~\ref{subsec:results_large_times}, we discuss the simplified emitter dynamics arising from Eq.~(\ref{eq:exp_series3}), which is only valid at times much larger than the photon round-trip time $d/c$. In Section~\ref{subsec:results_approximate}, we neglect delays due to the round-trip time and study the approximate emitter dynamics in the Markovian regime. Finally, in Sections~\ref{subsec:results_photon_1} and \ref{subsec:results_photon_2} we use Eq.~(\ref{eq:binomial_series44}) to calculate the spatial profile and the spectrum of the emitted photon wave packet, respectively.


\subsection{Exact dynamics of the quantum emitter}\label{subsec:results_exact}

In order to visualize the dynamics of the two-level quantum emitter given the initial state ${\ket{\psi(0)} = \ket{{\rm e},0}}$, we calculate the excitation probability
 \begin{equation}\label{eq:Pe_general}
    P_{\rm e}(t) = \left| \langle {\rm e},0 |\psi(t) \rangle \right|^2 . 
\end{equation}
Taking into account that the odd terms in $\ket{\psi(t)}$ do not contribute to the inner product, substituting the exact expression for $\ket{\psi_{\rm even}(t)}$ in Eq.~(\ref{eq:binomial_series4}) into the above equation and using the definition of $\Omega$ from Eq.~(\ref{condiextra2}) yields
\begin{align}\label{eq:Pe_exact}
\begin{split}
    P_{\rm e}(t) =  e^{- \Gamma t} \; \Biggl| \; \sum_{k=0}^{\infty}& \;\frac{a^k}{k!} \left( t- kd/c \right)^k \Theta \left( t-kd/c \right) \Biggr|^2 . 
\end{split}
\end{align}

Due to the step functions $\Theta(t-kd/c)$ in Eq.~(\ref{eq:Pe_exact}), we need to partition the time evolution into intervals of duration $d/c$. Physically, this is because there is a maximum number of round trips that a photon can complete up to a given time $t$, and this number of round trips is fixed within each interval. For example, for ${0 \leq t < d/c}$, only the ${k=0}$ term in Eq.~(\ref{eq:Pe_exact}) contributes, since no round trips could have occurred yet. This leads to
\begin{equation}\label{eq:free_space}
    P_{\rm e}(0 \leq t < d/c) = e^{-\Gamma t},
\end{equation}
which is the exponential decay expected for an emitter in an infinite waveguide with no mirror. Hence, the emitter is not affected by the mirror before the photon has had time to return from the mirror, as expected from causality~\cite{Dorner2002}. Note that, in the limit ${d \to \infty}$, the above condition always holds and the emitter decays exactly as in an infinite waveguide.

In the next time interval, ${d/c \leq t < 2d/c}$, only the ${k=0}$ and ${k=1}$ terms in Eq.~(\ref{eq:Pe_exact}) contribute, since a maximum of one round trip could have occurred up to this time. This leads to the modified evolution
\begin{equation}
    P_{\rm e}(d/c \leq t < 2d/c) = e^{-\Gamma t} \left| 1+a (t-d/c) \right|^2.
\end{equation}
More generally, for times $t$ with ${nd/c \leq t < (n+1)d/c}$, a photon can complete up to $n$ round trips, meaning that terms with ${k=0,1,\ldots,n}$ in Eq.~(\ref{eq:Pe_exact}) contribute. Hence,
\begin{equation} \label{n}
    P_{\rm e}(t) = e^{-\Gamma t} \; \left| \sum_{k=0}^n \frac{a^k}{k!} (t-kd/c)^k \right|^2
\end{equation}
when ${nd/c \leq t < (n+1)d/c}$. This equation provides an exact result for the dynamics of the waveguide-coupled emitter in the presence of the mirror.

\begin{figure}[t]
    \centering
    \includegraphics[width=\linewidth]{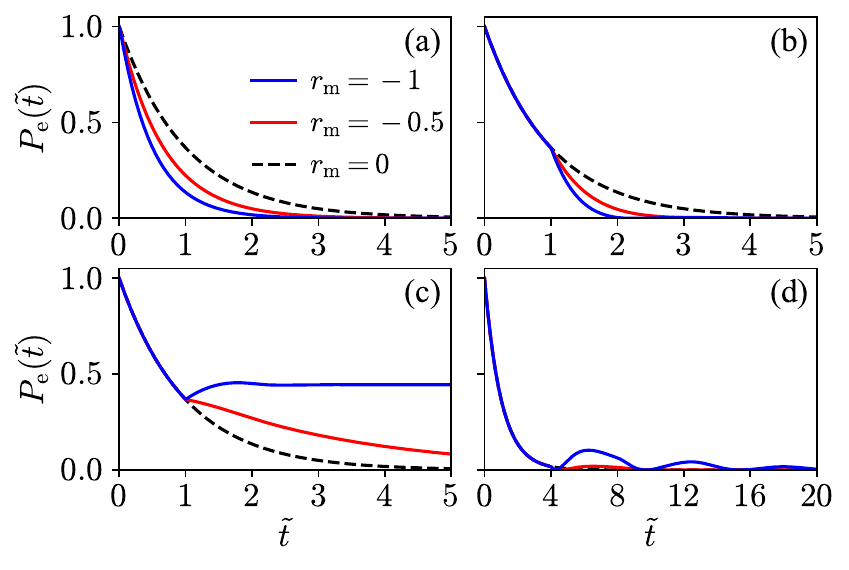}
    \caption{Two-level emitter excitation probability as a function of time (in normalized units), calculated using Eq.~(\ref{n}). Each panel shows results for three different reflection coefficients: ${r_{\rm m}=-1}$ (solid blue curves), ${r_{\rm m}=-0.5}$ (solid red curves), and ${r_{\rm m}=0}$ (dashed black curves). The round-trip time is (a) ${\tilde{\tau} = 0.01}$, (b), (c) ${\tilde{\tau} = 1}$, and (d) ${\tilde{\tau} = 4}$. The transition frequency of the emitter is chosen such that the corresponding round-trip phases are (a), (b), (d) ${\tilde{\omega}_{\rm e} \tilde{\tau} = \pi}$ and (c) ${\tilde{\omega}_{\rm e} \tilde{\tau} = 2\pi}$.}
    \label{fig:plots_non-Markovian}
\end{figure}

In Fig.~\ref{fig:plots_non-Markovian}, we visualize Eq.~(\ref{n}) for different sets of parameters. To do this, we use dimensionless units where all quantities are normalized to the decay rate $\Gamma$, i.e., ${\tilde{t} = t\Gamma}$, ${\tilde{\tau} = (d/c)\Gamma}$, and ${\tilde{\omega}_{\rm e} = \omega_{\rm e}/\Gamma}$. Each panel in the figure shows the time evolution in the case of a perfectly reflecting mirror (${r_{\rm m}=-1}$, solid blue curves), a semi-transparent mirror (${r_{\rm m}=-0.5}$, solid red curves), and a perfectly transparent mirror (${r_{\rm m}=0}$, dashed black curves). The last case is simply the infinite-waveguide result ${P_{\rm e}(\tilde{t}) = e^{-\tilde{t}}}$, which we obtain by setting ${a=0}$ in Eq.~(\ref{n}) (only the ${k=0}$ term contributes for all times).

Fig.~\ref{fig:plots_non-Markovian}(a) corresponds to the case where the round-trip time is significantly shorter than the emission lifetime (${\tilde{\tau} \ll 1}$). In this case, the decay of the emitter follows an exponential that is modified from the infinite-waveguide case by interference between emitted and reflected light when ${r_{\rm m} \neq 0}$. In Figs.~\ref{fig:plots_non-Markovian}(b) and (c), the round-trip time is comparable to the emission lifetime (${\tilde{\tau} = 1}$). In this case, we clearly see that the evolution follows an exponential decay until ${\tilde{t} = \tilde{\tau}}$, after which the emitter dynamics are modified. Depending on the round-trip phase ${\tilde{\omega}_{\rm e} \tilde{\tau} = \omega_{\rm e}d/c}$, the emission can be enhanced [Fig.~\ref{fig:plots_non-Markovian}(b)] or suppressed [Fig.~\ref{fig:plots_non-Markovian}(c)]. In particular, a perfect mirror can lead to population trapping where the excitation probability remains nonzero for all times~\cite{Regidor2021} [blue curve in Fig.~\ref{fig:plots_non-Markovian}(c)]. This is not possible with an imperfect mirror, where the emitter eventually decays to its ground state completely due to imperfect interference between emitted and reflected fields caused by photon leakage. 

In Fig.~\ref{fig:plots_non-Markovian}(d), the round-trip time is longer than the emission lifetime (${\tilde{\tau} = 4}$). Again, the emitter decays exponentially as in an infinite waveguide until ${\tilde{t} = \tilde{\tau}}$, after which it is partially re-excited by the returning photon. This happens every time the photon completes a round trip, but the amplitude of the re-excitation gradually reduces to zero because after each reabsorption the photon is equally likely to be emitted away from the mirror (i.e., out of the system) and towards the mirror. As expected, the re-excitation peaks are less pronounced when $|r_{\rm m}|$ is smaller.

As Fig.~\ref{fig:plots_non-Markovian} shows, the decay of the emitter can be very different in the presence of the mirror compared to having an infinite waveguide. In Appendix~\ref{app:trajectories} we compare the exact results of this subsection to those obtained with a numerical quantum trajectory approach and find excellent agreement between the two methods. This verifies that our model is consistent with more established techniques for modeling light--matter interactions in quantum optics, demonstrating the validity of our analytical method for modeling non-trivial dynamics in quantum-optical systems.


\subsection{Exponential decay for ${t \gg d/c}$}\label{subsec:results_large_times}

Next, we consider large times $t$ satisfying ${t \gg d/c}$ (or equivalently ${\tilde{t} \gg \tilde{\tau}}$). In this long-time limit, we can calculate the excitation probability $P_{\rm e}(t)$ using Eq.~(\ref{eq:exp_series3}). Substituting this result into Eq.~(\ref{eq:Pe_general}) yields
\begin{equation}
    P_{\rm e}(t) = \left| \xi_0 \; e^{-i \omega_e t} e^{-(\Gamma - 2\xi)t/2} \right|^2
     = |\xi_0|^2 \; e^{- (\Gamma - 2 {\rm Re}(\xi)) t},
\end{equation}
provided that we are in the convergence regime where we can calculate $\xi_0$ and $\xi$, as discussed below Eq.~(\ref{finalxi0}). The above equation describes an exponential decay with a modified decay rate 
\begin{equation}
\Gamma' = \Gamma - 2 {\rm Re}(\xi),
\end{equation}
which can be either less than or greater than the free-space decay rate $\Gamma$.

We illustrate this by considering the emitter and mirror parameters from Fig.~\ref{fig:plots_non-Markovian} (for the blue curves with ${r_{\rm m} = -1}$). For the parameters of Fig.~\ref{fig:plots_non-Markovian}(a) where $d/c$ is smallest, to a good approximation we can use the first order Maclaurin expansion $W_0(x) = x$ to show that $\xi_0 \approx 1/(1+ad/c) \approx 1$ and $\xi \approx a \approx -r_{\rm m}e^{i\omega_ed/c} \Gamma/2 = -0.5\Gamma$ [see Eqs.~(\ref{condiextra66}) and (\ref{finalxi0})]. This value of $\xi$ corresponds to a maximum effective decay rate ${\Gamma' = 2\Gamma}$, in agreement with the exact result in Fig.~\ref{fig:plots_non-Markovian}(a). In contrast, for the parameters of Figs.~\ref{fig:plots_non-Markovian}(b)-(d) where $d/c$ is significantly larger, $\xi_0$ diverges. A solution for $\xi$ exists in the case of Fig.~\ref{fig:plots_non-Markovian}(c), where ${\xi \approx 0.5\Gamma}$. This corresponds to the effective decay rate ${\Gamma' = 0}$, in agreement with Fig.~\ref{fig:plots_non-Markovian}(c) where the emission is suppressed after the initial transient time.


\subsection{Approximate exponential solution}\label{subsec:results_approximate}

If the round-trip time $d/c$ is small compared to other relevant timescales, such as the emission lifetime $1/\Gamma$, we can neglect it to a good approximation. Replacing ${t-kd/c}$ with $t$ in the exact result in Eq.~(\ref{eq:binomial_series4}) leads to
\begin{equation}
    \ket{\psi_{\rm even}(t)} = e^{-i\Omega t} \sum_{k=0}^{\infty} \frac{(at)^k}{k!}\ket{\rm{e},0} = e^{-i\Omega t} e^{at} \ket{\rm{e},0}.
\end{equation}
Using the definitions of $\Omega$ and $a$ in Eq.~(\ref{condiextra2}) and taking into account that ${e^{\Gamma d/2c} \approx 1}$ when ${d/c \ll 1/\Gamma}$, we obtain
\begin{equation}\label{eq:state_exp}
    \ket{\psi_{\rm even}(t)} = e^{-i\omega_e t} e^{-\Gamma t/2 \left( 1 + r_{\rm m} e^{i\omega_e d/c} \right)} \ket{\rm{e},0}.
\end{equation}
Here we have ignored the propagation time between the emitter and the mirror but kept the resulting round-trip phase factor $e^{i\omega_e d/c}$. In this Markovian regime, the excitation probability $P_{\rm e}(t)$ takes the simple form
\begin{equation}\label{eq:Pe_markovian}
    P_{\rm e}(t) = e^{-\Gamma t \left[ 1 + r_{\rm m}{\rm{cos}}(\omega_e d/c) \right]},
\end{equation}
which is an exponential decay modified by the interference arising from reflections at the mirror. Here we have assumed that $r_{\rm m}$ is real for simplicity. Note that making $r_{\rm m}$ complex would correspond to adding a constant phase to the round-trip phase, meaning that any result with complex $r_{\rm m}$ can be reproduced with a real $r_{\rm m}$ by adjusting the round-trip phase.

When ${r_{\rm m}\text{cos}(\omega_{\rm e}d/c) = 1}$, ${P_{\rm e}(t) = e^{-2\Gamma t}}$, so the decay rate is doubled from $\Gamma$ to $2\Gamma$. A similar effect known as superradiance occurs in the emission from two quantum emitters with dipole-dipole interactions~\cite{Dicke1954, Scheibner2007, Tiranov2023}. In our case, the emitter effectively interacts with its mirror image and its emission is enhanced due to constructive interference between the emitted and reflected fields. On the other hand, when ${r_{\rm m}\text{cos}(\omega_{\rm e}d/c) = -1}$, ${P_{\rm e}(t) = 1}$ for all times and the emission is completely suppressed. This is analogous to subradiance, and is caused by destructive interference between emitted and reflected light. Note that the special cases of maximum enhancement and complete suppression of emission can only be observed with a perfect mirror where ${|r_{\rm m}| = 1}$. When ${|r_{\rm m}| < 1}$, the decay rate of the emitter is somewhere between the two extreme values. This is in agreement with the results in the previous subsections where we observed the same modifications for the same parameters, but only after the round-trip time $d/c$.

\begin{figure}[t]
    \centering
    \includegraphics[width=\linewidth]{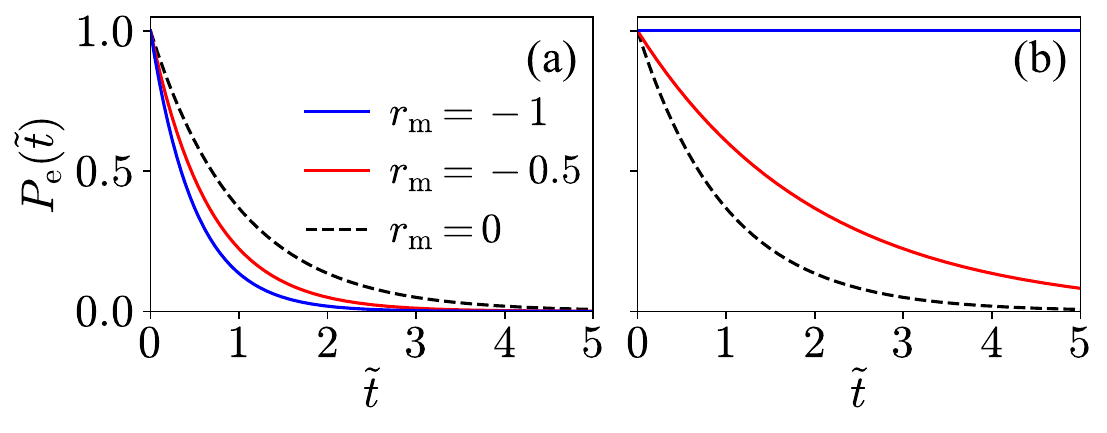}
    \caption{Approximate two-level emitter excitation probability in the Markovian regime [Eq.~(\ref{eq:Pe_markovian})] as a function of time. As in Fig.~\ref{fig:plots_non-Markovian}, we consider the reflection coefficients ${r_{\rm m}=-1}$ (solid blue curves), ${r_{\rm m}=-0.5}$ (solid red curves), and ${r_{\rm m}=0}$ (dashed black curves). The round-trip time is ${\tilde{\tau} = 0.01}$, and the round-trip phases are (a) ${\tilde{\omega}_{\rm e} \tilde{\tau} = \pi}$ and (b) ${\tilde{\omega}_{\rm e} \tilde{\tau} = 2\pi}$.}
    \label{fig:plots_markovian}
\end{figure}

Figure~\ref{fig:plots_markovian} shows $P_{\rm e}(t)$ in Eq.~(\ref{eq:Pe_markovian}) for two different round-trip phases: ${\tilde{\omega}_{\rm e} \tilde{\tau} = \pi}$ [Fig.~\ref{fig:plots_markovian}(a)] and ${\tilde{\omega}_{\rm e} \tilde{\tau} = 2\pi}$ [Fig.~\ref{fig:plots_markovian}(b)], with the chosen round-trip time ${\tilde{\tau} = 0.01}$ ensuring that the Markovian approximation is valid. As expected, Fig.~\ref{fig:plots_markovian}(a) is identical to Fig.~\ref{fig:plots_non-Markovian}(a), as the approximate result coincides with the exact solution in the limit of small round-trip times. For the solid blue curve in Fig.~\ref{fig:plots_markovian}(a), ${r_{\rm m} \text{cos}(\omega_{\rm e}d/c) = 1}$, so the emission rate is maximally enhanced. For the solid red curve the reflection is reduced, so the emission rate is closer to that expected in an infinite waveguide with no mirror (given by the dashed black curve).
In Fig.~\ref{fig:plots_markovian}(b), the round-trip phase is the same as in Fig.~\ref{fig:plots_non-Markovian}(c), but the round-trip time is significantly shorter. For the solid blue curve in Fig.~\ref{fig:plots_markovian}(b), ${r_{\rm m}\text{cos}(\omega_{\rm e}d/c) = -1}$, so the emission is completely suppressed. This behavior is similar to the blue curve in Fig.~\ref{fig:plots_non-Markovian}(c), but in that case the emission only becomes suppressed after the round-trip time, resulting in a steady-state emitter excitation probability below unity due to the initial emission process. Again, for the solid red curve where reflection is reduced, the emission rate is closer to that in an infinite waveguide with no mirror.

\begin{figure}[t]
    \centering
    \includegraphics[width=\linewidth]{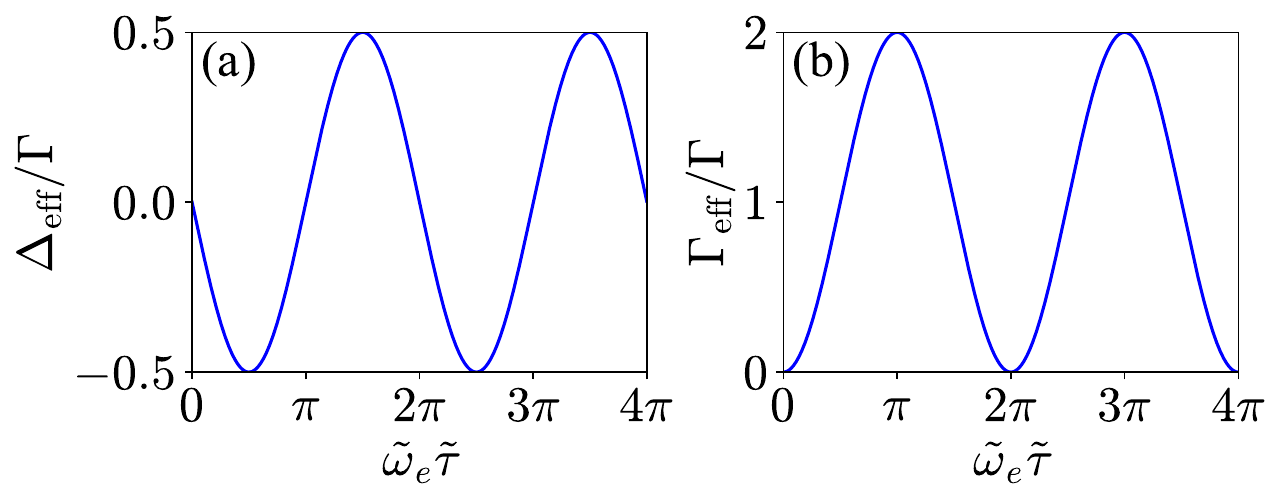}
    \caption{(a) Transition frequency shift $\Delta_{\rm eff}$ and (b) effective decay rate $\Gamma_{\rm eff}$ as a function of the round-trip phase ${\tilde{\omega}_e \tilde{\tau} = \omega_e d/c}$, calculated using Eqs.~(\ref{eq:delta}) and (\ref{eq:gamma}), respectively. The reflection coefficient of the mirror is ${r_{\rm m} = -1}$.}
    \label{fig:shift_decay_rate}
\end{figure}

To look at the effect of the mirror in the Markovian regime in more detail, we express the state in Eq.~(\ref{eq:state_exp}) in the form
\begin{equation}\label{eq:exp_series_v2}
    \ket{\psi_{\rm even}(t)} = e^{-i(\omega_e + \Delta_{\rm eff})t} e^{-\Gamma_{\rm eff}t/2} \ket{\rm{e},0},
\end{equation}
where
\begin{equation}\label{eq:delta}
    \Delta_{\rm eff} = \left(\frac{\Gamma}{2}\right) r_{\rm m} {\rm{sin}}\left(\frac{\omega_e d}{c}\right)
\end{equation}
and
\begin{equation}\label{eq:gamma}
    \Gamma_{\rm eff} = \Gamma \left[1 + r_{\rm m} {\rm{cos}}\left(\frac{\omega_e d}{c}\right) \right]
\end{equation}
denote the effective shift in the transition frequency and the effective decay rate of the emitter, respectively. In this dressed-atom picture, the presence of the mirror Hamiltonian $H_{\rm m}$ in Eq.~(\ref{eq:H_m}) modifies the bare transition frequency $\omega_e$ and decay rate $\Gamma$ of the emitter. Note that, when ${r_{\rm m} = 0}$, ${\Delta_{\rm eff} = 0}$ and ${\Gamma_{\rm eff} = \Gamma}$, so Eq.~(\ref{eq:exp_series_v2}) reduces to
\begin{equation}
    \ket{\psi_{\rm even}(t)} = e^{-i\omega_et} e^{-\Gamma t/2} \ket{e,0},
\end{equation}
which is the expected exponential decay without a mirror~\cite{Hartwell2026}. However, in general both $\Delta_{\rm eff}$ and $\Gamma_{\rm eff}$ have an oscillatory dependence on the round-trip phase ${\omega_e d/c = \tilde{\omega}_e \tilde{\tau}}$, as shown in Fig.~\ref{fig:shift_decay_rate}. This observation is consistent with the results of previous experiments involving spontaneous emission near mirror surfaces~\cite{Drexhage1970, Amos1997, Eschner2001, Frimmer2013, Qvotrup2025} (the oscillations persist for all round-trip phases due to the Markovian approximation, but in the non-Markovian regime $\Delta_{\rm eff}$ and $\Gamma_{\rm eff}$ would approach the free-space values in the limit ${d \rightarrow \infty}$). The amplitude of the oscillations is determined by the reflection coefficient $r_{\rm m}$. In the figure, we assume a perfect mirror with ${r_{\rm m} = -1}$, in which case the shift in transition frequency oscillates between the two values ${\Delta_{\rm eff} = \pm \Gamma/2}$, and the decay rate oscillates between ${\Gamma_{\rm eff} = 0}$ (maximal suppression) and ${\Gamma_{\rm eff} = 2\Gamma}$ (maximal enhancement of emission).

We can explain the observed emission enhancement and suppression in the following way. Consider a perfect mirror with ${r_{\rm m}=-1}$. In this case, complete suppression of emission occurs when the round-trip phase is an integer multiple of $2\pi$: ${\omega_{\rm e}d/c =  2m\pi}$ for any integer $m$. This occurs when the separation between the emitter and the mirror is a half-integer multiple of the transition wavelength (${d/2 = m\lambda_{\rm e}/2}$, with ${\lambda_{\rm e} = 2\pi c/\omega_{\rm e}}$), i.e., the emitter is located at a field node. On the other hand, maximum enhancement occurs when the round-trip phase is an odd integer multiple of $\pi$: ${\omega_{\rm e}d/c =  m\pi}$ with odd $m$. This occurs when the emitter--mirror separation is ${d/2 = m\lambda_{\rm e}/4}$ ($= \lambda_{\rm e}/4, 3\lambda_{\rm e}/4, \ldots$), i.e., the emitter is located at a field antinode. The fact that the emission rate depends strongly on the round-trip phase $\omega_{\rm e} d/c$ means that we can control the decay rate of the emitter either by tuning the transition frequency $\omega_{\rm e}$ or by adjusting the emitter--mirror round-trip time (by changing the physical distance $d$ or the speed of light $c$ in the waveguide).


\subsection{Spatial profile of the photon wave packet}\label{subsec:results_photon_1}

In the remainder of this section, we consider the odd terms $\ket{\psi_{\rm odd}(t)}$ of the full state of the system to study the properties of the generated photon wave packet. In particular, we focus on the part of the photon state where ${x<0}$, corresponding to the photon leaving the system to the left of the emitter (see Fig.~\ref{fig:diagram}). From Eq.~(\ref{eq:binomial_series44}), it follows that the part of the state $\ket{\psi_{\rm odd}(t)}$ where a field excitation is present at ${x<0}$ is given by
\begin{align}\label{eq:state_left}
\begin{split}
    \ket{\psi_{\rm L}(t)} &= \; -ig \sum_{k=0}^{\infty} \int_0^t {\rm d}t_1 \;  e^{-i \Omega t_1} \frac{a^k}{k!} \; \Theta \left( t_1 - kd/c \right)  \\
    & \;\times \left[ r_{\rm m} \; \Theta\left( c(t-t_1) - d \right) \ket{{\rm g}, 1_{-1}(d - c(t-t_1)} \right.\\
     & \; \left. + \ket{{\rm g}, 1_{-1}(-c(t-t_1))} \right]  \left( t_1- kd/c \right)^{k}.
\end{split}
\end{align}
This state contains two contributions---one where the photon is emitted directly to the left and one where the photon leaves the system to the left after being reflected. These contributions interfere with each other, resulting in a wave packet that is modified from the simple exponential profile that is obtained in an infinite waveguide or in free space~\cite{Hartwell2026}. The step function ${\Theta( c(t-t_1) - d)}$ ensures that we only have the part of the reflected field that has passed the emitter and has reached the ${x<0}$ region.

\begin{figure}[t]
    \centering
    \includegraphics[width=\linewidth]{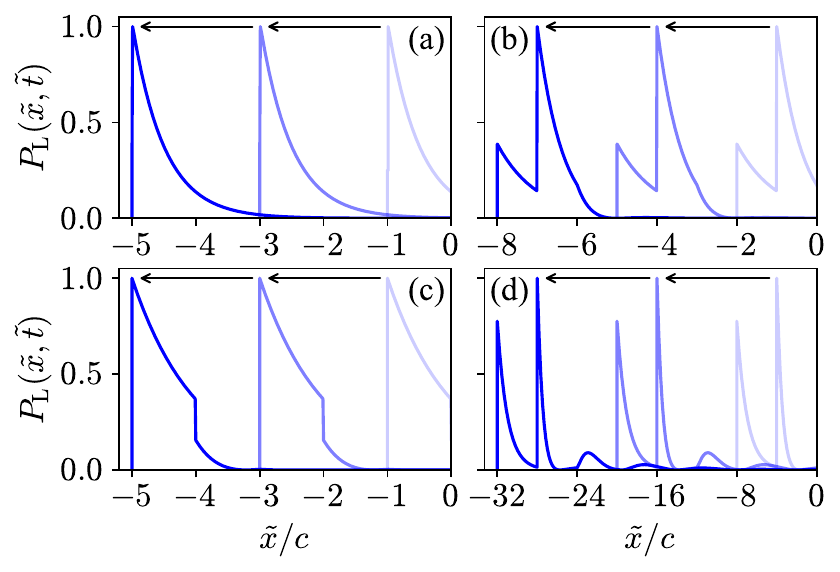}
    \caption{Photon wave packet emitted in the $-x$ direction (to the left, away from the mirror) as a function of normalized position $\tilde{x}/c$ [calculated using Eq.~(\ref{eq:P_L})]. The parameters used to obtain these wave packets are the same as those used for the blue curves in Fig.~\ref{fig:plots_non-Markovian} (i.e., where ${r_{\rm m}=-1}$). In each subfigure, we show the wave packet at three different times ${\tilde{t}}$ as it moves away from the emitter located at ${x=0}$ (in the direction indicated by the black arrows). All the wave packets are rescaled to have a peak value of one.}
    \label{fig:wave_packets}
\end{figure}

The derivation of the probability density $P_{\rm L}(x,t)$ for finding a photon at time $t$ at a position $x$ with ${x<0}$ can be found in Appendix~\ref{app:position}. Using Eq.~(\ref{eq:state_left}), we show that this probability density can be written as 
\begin{equation}
    P_{\rm L}(x,t) = \left| \Phi_{\rm L}(x,t) \right|^2 ,
\end{equation}
where $\Phi_{\rm L}(x,t)$ is the amplitude of the $\ket{{\rm g}, 1_{-1}(x)}$ state [after performing a transformation of Eq.~(\ref{eq:state_left}) from time $t_1$ to position $x$], and equals
\begin{equation}\label{eq:phi_x}
    \Phi_{\rm L}(x,t) = -\frac{ig}{c} e^{-i \Omega (x/c+t)} \phi_{\rm L}(x,t) .
\end{equation}
The function $\phi_{\rm L}(x,t)$ is given in Eq.~(\ref{eq:amplitude_x}) for the interval ${-ct+nd \leq x < -ct+(n+1)d}$ with ${n \geq 1}$ (in the first interval where ${n=0}$, ${\phi_{\rm L}(x,t) = 1}$). We therefore have
\begin{equation}\label{eq:P_L}
    P_{\rm L}(x,t) = \frac{|g|^2}{c^2} e^{-\Gamma (x/c+t)} \left|\phi_{\rm L}(x,t)\right|^2,
\end{equation}
using ${\Omega = \omega_e - i\Gamma/2}$.

In Fig.~\ref{fig:wave_packets}, we visualize Eq.~(\ref{eq:P_L}) to show the photon wave packet corresponding to each panel in Fig.~\ref{fig:plots_non-Markovian} (with ${r_{\rm m}=-1}$), at different times $\tilde{t}$ as a function of the normalized position ${\tilde{x}/c = (x/c)\Gamma}$. In Fig.~\ref{fig:wave_packets}(a), the wave packet has an exponential profile, which mirrors the exponential decay of the emitter in Fig.~\ref{fig:plots_non-Markovian}(a), where the round-trip time is negligible and the Markovian approximation is valid. In Fig.~\ref{fig:wave_packets}(b), we observe constructive interference between the field component emitted to the left and the component reflected from the mirror, resulting in a second, enhanced peak in the spatial profile [this corresponds to the enhanced emission in Fig.~\ref{fig:plots_non-Markovian}(b)]. Conversely, Fig.~\ref{fig:wave_packets}(c) demonstrates destructive interference between emitted and reflected field components, resulting in a sudden decrease in the amplitude of the total wave packet moving to the left [this corresponds to the suppressed emission in Fig.~\ref{fig:plots_non-Markovian}(c)]. In Fig.~\ref{fig:wave_packets}(d), the increased round-trip time reduces the overlap between emitted and reflected light, giving rise to two clear peaks in the spatial profile. These peaks are followed by small oscillations arising from partially re-exciting the emitter after every round trip, which gradually decay to zero [see Fig.~\ref{fig:plots_non-Markovian}(d)].


\subsection{Spectral profile of the photon wave packet} \label{subsec:results_photon_2}

Using our analytical results, we can also study the spectral profile of the emitted light. We do this by considering a large fixed time ${t \gg 1/\Gamma}$ such that the emitter has decayed to its ground state, and taking the Fourier transform of the amplitude $\Phi_{\rm L}(x,t)$ in Eq.~(\ref{eq:phi_x}), converting the time variable $x/c$ to a frequency variable $\omega$. This leads to the amplitude $\tilde{\Phi}_{\rm L}(\omega)$ in the frequency domain, which we use to calculate the spectral probability density
\begin{equation}\label{eq:P_L_freq}
    P_{\rm L}(\omega) = \left| \tilde{\Phi}_{\rm L}(\omega) \right|^2.
\end{equation}

\begin{figure}[t]
    \centering
    \includegraphics[width=\linewidth]{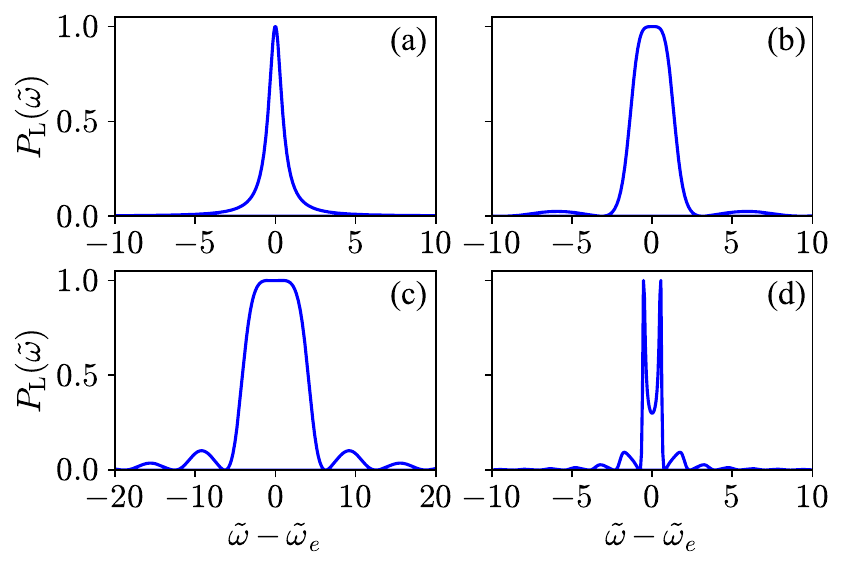}
    \caption{Spectral envelope of the emitted photon wave packet, calculated in the long-time limit using Eq.~(\ref{eq:P_L_freq}) (once the emitter returns to its ground state). In (a) ${r_{\rm m}=0}$ while in (b)-(d) ${r_{\rm m} = -1}$, with other parameters being the same as in Fig.~\ref{fig:wave_packets}(b)-(d). The Fourier transform from time $x/c$ to frequency $\omega$ was computed numerically in Python using the \texttt{fftpack.fft} function from the scipy library.}
    \label{fig:spectra}
\end{figure}

Figure~\ref{fig:spectra} shows examples of spectral profiles as a function of the normalized frequency ${\tilde{\omega} = \omega/\Gamma}$. Figure~\ref{fig:spectra}(a) corresponds to ${r_{\rm m} = 0}$, which produces the expected Lorentzian spectrum as for a two-level emitter in an infinite waveguide with no mirror (the Fourier transform of an exponential profile)~\cite{Hartwell2026}. Figures~\ref{fig:spectra}(b)-(d) show the spectra corresponding to the spatial profiles in Figs.~\ref{fig:wave_packets}(b)-(d), demonstrating the effect of the interference between the emitted and reflected field components on the emission spectrum in the regime where the round-trip time $d/c$ is significant and the decay is non-exponential.


\section{Conclusion}\label{sec:conc}

In this paper, we describe the dynamics of a two-level quantum emitter in a one-dimensional waveguide with a partially transparent mirror using a Schr\"odinger equation based on a locally-acting Hamiltonian. We calculate the evolution of an initially-excited emitter by using a Dyson series expansion to solve the Schr\"odinger equation. Our calculations show the effect of interference between emitted and reflected field excitations on the dynamics of the emitter and the field in the waveguide. For example, the time evolution of the probability $P_{\rm e}(t)$ of finding the emitter in its excited state is in general non-exponential and only becomes exponential when the photon round-trip time is negligible or after a sufficiently long transient time (see Figs.~\ref{fig:plots_non-Markovian} and \ref{fig:plots_markovian}). In the Markovian regime where the emitter--mirror round-trip time is neglected, the dynamics can be described using a dressed-atom picture where the transition frequency and decay rate are modified by the mirror (see Fig.~\ref{fig:shift_decay_rate}). The fact that the effective decay rate $\Gamma_{\rm eff}$ depends strongly on the round-trip phase $\omega_{\rm e} d/c$ means that we can control the decay rate of the emitter either by tuning the transition frequency or by adjusting the emitter--mirror round-trip time. On the other hand, in the regime where the photon round-trip time between the emitter and the mirror is significant compared to the emission lifetime, the presence of the mirror can significantly alter the emitter dynamics as well as the spatial and spectral profiles of the generated photon wave packet (see Figs.~\ref{fig:wave_packets} and \ref{fig:spectra}).

The emitter--mirror system that we study here can be realized experimentally using a quantum emitter inside a one-dimensional waveguide terminated at one end. Quantum nanophotonic devices typically operate in the Markovian regime, where the emitter decays exponentially with a rate modified by the mirror~\cite{Qvotrup2025}. This is due to the small sizes of these devices, together with short photon propagation times in comparison to the emission lifetime. For example, in a typical quantum dot--waveguide system, ${d \sim 10}$~$\upmu$m and $1/\Gamma \sim 1$~ns~\cite{Qvotrup2025, Laucht2012, Arcari2014}, so the photon round-trip time ${d/c \sim 0.1}$~ps (with ${c\sim10^8}$~m/s) would be orders of magnitude shorter than the emission lifetime (${d/c \ll 1/\Gamma}$). In this regime, the exponential decay can be controlled with the round-trip phase $\omega_{\rm e}d/c$. Since the position of integrated solid-state emitters such as quantum dots is fixed, the phase can be adjusted either by tuning the transition frequency $\omega_{\rm e}$ (e.g., electrically~\cite{Hallett2018}) or by changing the speed of light $c$. The latter can be achieved using a photonic crystal waveguide and deforming the lattice (e.g., by applying strain), which changes the group index and hence the group velocity. This control can potentially be used to construct a waveguide-integrated single-photon source with a tunable emission rate.

In order to observe non-Markovian effects such as non-exponential decay, we require the emitter--mirror round-trip time to be comparable to the emission lifetime [${\tilde{\tau} = (d/c)\Gamma \sim 1}$]. In the quantum dot--waveguide system mentioned above, this would require either significantly shortening the lifetime of the emitter or increasing the photon propagation time, e.g., using the slow-light effect in photonic crystal waveguides~\cite{Krauss2007, Arcari2014, Siampour2023}. Alternatively, the non-Markovian effects can be observed in circuit quantum acoustodynamics, where artificial atoms are coupled to phonons instead of photons~\cite{Gustafsson2014, Manenti2017}. Phonons are orders of magnitude slower than photons and hence make the non-Markovian regime more accessible. Experimentally, non-Markovian evolution has been observed using a “giant atom" architecture with superconducting circuits, where an artificial atom is coupled to a waveguide at two different locations~\cite{Andersson2019, Xiao2025} (this configuration can be mapped onto an emitter in front of a mirror~\cite{Kockum2021}). Since a non-exponential decay translates into a non-exponential wave packet (see Fig.~\ref{fig:wave_packets}), this system may find applications in pulse-shaping experiments, where the time delay is chosen based on a desired wave packet shape.

Finally, note that our model neglects imperfections like photon loss into non-guided modes and decoherence in the quantum emitter, which are inevitable in solid-state emitters due to interactions with the environment. Ignoring decoherence is reasonable provided that the optical coherence time is significantly longer than the emitter--mirror round-trip time, which allows the necessary interference to occur before the emitter decoheres. Note that close to lifetime-limited coherence times on the order of nanoseconds have been measured in quantum dots~\cite{Zhai2020}, while superconducting qubits have shown coherence times reaching milliseconds~\cite{Place2021, Somoroff2023, Bland2025}. In addition, emitter--waveguide coupling with near-unity efficiency has been achieved in these systems~\cite{Arcari2014, Scarpelli2019, Sheremet2023}. In the presence of imperfect coupling, we expect that photon loss would inhibit the delay-feedback-induced interference effects if the loss rate becomes comparable to the emitter--waveguide coupling rate. Our model could be adapted to account for photon loss by including additional terms in the local interaction Hamiltonian [Eq.~(\ref{eq:H_int})], which describe coupling to a loss channel (effectively a second waveguide with no mirror, such that lost excitations never return). We leave this extension as a possible avenue for future work.


\section*{Acknowledgments}

This work was supported by the Engineering and Physical Sciences Research Council [grant numbers EP/W524360/1, EP/W524372/1, EP/Z533208/1, and 2885425].





\appendix


\begin{widetext}

\section{Calculation of the coefficients $c_n(t)$ in Eq.~(\ref{E30})}\label{AppA}

To obtain the coefficient $c_4(t)$ from $c_2(t)$ using Eq.~(\ref{eq:c_n+2}), the following two integrals need to be used:
\begin{align}\label{eq:c_n+2x}
\begin{split}
\int_0^t {\rm d} t_1 \; c_2(t_1) 
=& -\frac{|g|^2}{2c} \left[ t^2 +  r_{\rm m}  e^{i\omega_{\rm e} d/c} \; \Theta \left( t-d/c \right) \left( t-d/c \right)^2 \right ] , \\
\int_0^{t-d/c} \; {\rm d} t_1 \; c_2(t_1) 
=& -\frac{|g|^2}{2c} \left[ \left( t - d/c \right)^2 +  r_{\rm m}e^{i\omega_{\rm e} d/c} \; \Theta \left( t-2d/c \right) \left( t-2d/c \right)^2 \right ] .
\end{split}
\end{align}
Substituting these into Eq.~(\ref{eq:c_n+2}) with ${n=2}$ leads to
\begin{equation}
\label{eq:U_4_loops}
c_4(t) = \; \frac{|g|^4}{2c^2} \left[ t^2 + 2r_{\rm m}e^{i\omega_{\rm e} d/c} \; \Theta \left( t-d/c \right) \left( t- d/c \right)^2 + r_{\rm m}^2 e^{2i \omega_{\rm e} d/c} \; \Theta \left( t-2d/c \right) \left( t- 2d/c \right)^2 \right] .
\end{equation}
In the same way that the coefficient $c_2(t)$ contains contributions from the single-loop reabsorption processes shown in Fig.~\ref{fig:one_loop}, $c_4(t)$ contains contributions from the two-loop processes shown in Fig.~\ref{fig:two_loops}. Here, the interaction Hamiltonian $H_1$ is applied four times, which corresponds to two emission--absorption cycles. As Fig.~\ref{fig:two_loops} shows, there are four contributions to $c_4(t)$, as each of the two reabsorption processes can happen instantly or after the time delay $d/c$. The two contributions where there is one instant reabsorption and one delayed reabsorption are identical (there are two possible orderings in this case).

\begin{figure}[h]
    \centering
    \includegraphics[width=0.5\textwidth]{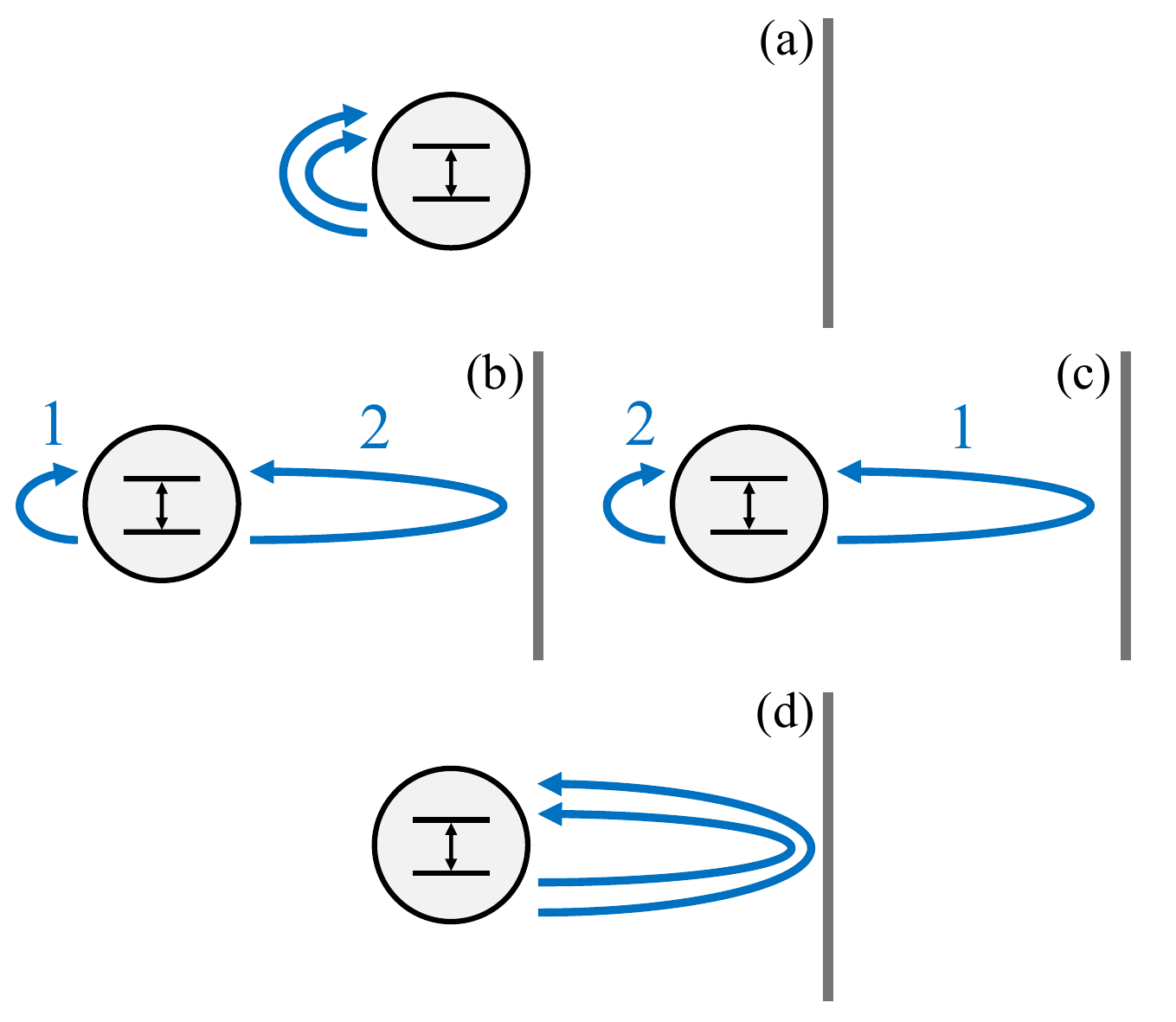}
    \caption{Reabsorption processes contributing to the Dyson series term $U_4(t,0)\ket{{\rm e},0}$. (a) Two instant reabsorption processes, corresponding to the first term in Eq.~(\ref{eq:U_4_loops}). (b) Instant reabsorption followed by delayed reabsorption, and (c) delayed reabsorption followed by instant reabsorption. These processes lead to the same contribution, providing the second term in Eq.~(\ref{eq:U_4_loops}). (d) Two delayed reabsorption processes where the photon completes two round trips between the emitter and the mirror, leading to the last term in Eq.~(\ref{eq:U_4_loops}).}
    \label{fig:two_loops}
\end{figure}

Analogously, one can substitute Eq.~(\ref{eq:U_4_loops}) into Eq.~(\ref{eq:c_n+2}) with ${n=4}$ to show that
\begin{align}\label{eq:U_6_loops}
\begin{split}
c_6(t) = -\frac{|g|^6}{3! \; c^3}&\left[ t^3 + 3r_{\rm m} e^{i\omega_{\rm e} d/c} \; \Theta \left( t-d/c \right) \left( t- d/c \right)^3 
+ 3 r_{\rm m}^2 e^{2i \omega_{\rm e} d/c} \; \Theta \left( t-2d/c \right) \left( t- 2d/c \right)^3\right. \\
&\left. \; + r_{\rm m}^3 e^{3i \omega_{\rm e} d/c} \; \Theta \left( t-3d/c \right) \left( t- 3d/c \right)^3 \right].
\end{split}
\end{align}
The coefficient $c_6(t)$ contains all the contributions from three reabsorption processes, where each one can happen instantly or after the delay $d/c$. In Eqs.~(\ref{eq:U_4_loops}) and (\ref{eq:U_6_loops}) we clearly see the binomial coefficients appearing, which arises from the different orderings of the instantaneous and delayed reabsorption processes in each contribution. More generally, there are ${\binom{n}{k} = n!/k!(n-k)!}$ ways of arranging $k$ instantaneous and $n-k$ delayed reabsorption processes among a total of $n$ reabsorption processes (this is illustrated for ${n=1}$ in Fig.~\ref{fig:one_loop} and for ${n=2}$ in Fig.~\ref{fig:two_loops}). Continuing the derivation of the remaining higher order terms in the Dyson series for even $n$, we end up with the expression for $c_n(t)$ in Eq.~(\ref{eq:binomial_series}).


\section{Calculation of $\ket{\psi_{\rm even}(t)}$ in terms of the Lambert $W$ function}\label{app:lambert_W}

Here we demonstrate how to write the analytic expression for the state $\ket{\psi_{\rm even}(t)}$ in Eq.~(\ref{eq:binomial_series4}) in terms of the Lambert $W$ function, as given in Eq.~(\ref{W1}). The complex phase $e^{-i\Omega t}$ will be kept as an overall phase of the state and not included as part of the state amplitude in the following calculations. In this derivation, we shall first determine the Laplace transform of the state amplitude given in Eq.~(\ref{eq:binomial_series4}) and then carry out the inverse Laplace transform, which will give us the final result. The Laplace transform $F(s)$ of the amplitude in Eq.~(\ref{eq:binomial_series4}) is given by
\begin{equation}
\label{Laplace1}
F(s) = \int_0^\infty{\rm d}t\;\sum_{k=0}^\infty\frac{a^k}{k!}\,\theta(t-kd/c)(t-kd/c)^k\,e^{-st} \, .
\end{equation}
By bringing the time integration inside the summation, the step function acts to raise the lower limit of the integral from zero to $kd/c$. The substitution $t \to t + kd/c$ then returns the lower integration limit to zero and gives
\begin{equation}
\label{Laplace2}
F(s) = \sum_{k=0}^\infty\frac{a^k}{k!}\int_0^\infty{\rm d}t\;t^ke^{-s(t+kd/c)}\, .
\end{equation}
By taking the time-independent part of the exponential out of the integral, we see that we are left with the Laplace transform of $t^k$ under a sum over all natural numbers, including zero. Using the known Laplace transform of a polynomial, we then find that
\begin{equation}
\label{Laplace3}
F(s) = \sum_{k=0}^\infty\frac{a^k}{k!}e^{-skd/c}\left(\frac{k!}{s^{k+1}}\right) = \frac{1}{s}\sum_{k=0}^\infty\left[(a/s)\,e^{-sd/c}\right]^k
\end{equation}
and, assuming convergence, the sum of the geometric series can be carried out. The final result is
\begin{equation}
\label{Laplace4}
F(s) = \frac{1}{s-ae^{-sd/c}}\, .
\end{equation}
Note here that $F(s)$ diverges when the denominator approaches zero, which is equivalent to when $(a/s)e^{-sd/c}$ in the geometric series approaches unity.

The next step of the calculation is to perform the inverse Laplace transformation to give us back the excited state amplitude, which we shall call $f(t)$. To carry out the inversion, we use Bromwich's inversion formula, which involves complex integration of $F(s)$ along an axis with a constant real value chosen such that all poles of the Laplace transform lie to the left of this axis. In this problem, the Laplace transform (\ref{Laplace4}) has poles when $s = ae^{-sd/c}$ or equivalently when $(sd/c)e^{sd/c} = ad/c$. The solutions to this equation are related to the branches of the Lambert $W$ function: $s_k = (c/d)\,W_k(ad/c)$. As there is a countably infinite number of solutions $W_k$ (with integer $k$), there is an infinite number of poles. Rather than solving the integration explicitly, by closing the contour to the left using a semicircular path of infinite radius for $\theta \in [\pi/2, 3\pi/2]$ in the complex plane, we can use Cauchy's residue theorem to find $f(t)$. In particular,
\begin{equation}
\label{Laplace5}
f(t) = \lim_{s \to s_k}\sum_k\,(s-s_k)\frac{e^{st}}{s-ae^{-sd/c}} = \sum_{k}\,\frac{e^{s_kt}}{1+(ad/c)e^{-s_kd/c}},
\end{equation}
where the sum is performed over all poles. By inserting the branches of the Lambert $W$ function as the poles, deduced above, we can then show that the excited state amplitude is given by
\begin{equation}
\label{Laplace6}
f(t) = \sum_{k = -\infty}^{\infty}\frac{e^{[W_k(ad/c)]ct/d}}{1+(ad/c)e^{-W_k(ad/c)}}.
\end{equation}
To simplify once more, we make use of the definition of the Lambert $W$ function, $W_k(ad/c)\,e^{W_k(ad/c)} = ad/c$, which can be substituted into the denominator of Eq.~(\ref{Laplace6}). This finally leads to 
\begin{equation}
\label{Laplace7}
f(t) = \sum_{k = -\infty}^{\infty}\frac{e^{[W_k(ad/c)]ct/d}}{1+W_k(ad/c)},
\end{equation}
which gives the amplitude in Eq.~(\ref{W1}) in the main text.


\section{Calculation of $\ket{\psi_{\rm even}(t)}$ for large times $t$}\label{AppB}

To derive a simpler expression for $\ket{\psi_{\rm even}(t)} $ from the general result in Eq.~(\ref{eq:binomial_series4}), we assume that ${t \gg d/c}$ and set ${\Theta(t-kd/c)=1}$. We can then express this state as 
\begin{equation} \label{condiextra6}
\ket{\psi_{\rm even}(t)} = e^{-i \Omega t} \, f(t) \; \ket{{\rm e},0},
\end{equation}
with the function $f(t)$ given by
\begin{align} \label{eq:binomial_series4extra}
\begin{split}
f(t) =& \; \sum_{k=0}^{\infty} \frac{a^k}{k!} \left( t- kd/c \right)^{k},
\end{split}
\end{align}
where $a$ is given in Eq.~(\ref{condiextra2}). If we take the time derivative of $f(t)$, we obtain
\begin{align} \label{condiextra4}
\begin{split}
f'(t) = & \; \sum_{k=1}^{\infty} \frac{a^k}{(k-1)!} \left( t- kd/c \right)^{k-1} \\
= & \; \sum_{k'=0}^{\infty} \frac{a^{k'+1}}{k'!} \left( t- d/c - k'd/c \right)^{k'} \\
= & \; a \; f(t-d/c),
\end{split}
\end{align}
where we used the substitution ${k = k'+1}$ to go from the first line to the second line. Solving this differential equation reveals that 
\begin{equation} \label{condiextra5}
f(t) = \xi_0 \; e^{\xi t}
\end{equation}
with $\xi_0 = f(0)$ given in Eq.~(\ref{condiextra3}) and $\xi$ satisfying Eq.~(\ref{condiextra66}). This solution can be verified by taking the time derivative and comparing the result to the right-hand side of Eq.~(\ref{condiextra4}), which leads to Eq.~(\ref{condiextra66}). Substituting $f(t)$ in Eq.~(\ref{condiextra5}) back into Eq.~(\ref{condiextra6}) yields Eq.~(\ref{eq:exp_series3}).


\section{Quantum trajectory model and comparison of results}\label{app:trajectories}

Here we provide details of the numerical quantum trajectory model that we used to test the validity of our analytical approach. In Section~\ref{app_subsec:model} we construct the trajectory model by following the methodology presented in Ref.~\cite{Regidor2021} and adapting their model to account for an imperfect mirror (see also Ref.~\cite{Martin2025}). In Section~\ref{app_subsec:comparison}, we provide a comparison between the results of our model and the results of the quantum trajectory model.


\subsection{Model}\label{app_subsec:model}

In the quantum trajectory model, we discretize the waveguide region of interest into a series of spatial boxes of width $\Delta t$ in time, as shown in Fig.~\ref{fig:trajectory_diagram}. This allows us to simulate the time evolution in discrete time steps $\Delta t$. We start from the $k$-space Hamiltonian for a two-level emitter coupled to an infinite one-dimensional waveguide:
\begin{align}\label{eq:H_k}
\begin{split}
    H =&\; \omega_{\rm e} \sigma^+ \sigma^- + \int_0^{\infty} \omega(k) a_R^{\dagger}(k)a_R(k) dk + \int_{-\infty}^0 \omega(k) a_L^{\dagger}(k)a_L(k) dk\\[0.05in]
    &+ \int_0^{\infty} \left[ \sqrt{\frac{V_R}{2\pi}} a_R^{\dagger}(k) \sigma^- + \sqrt{\frac{V_R^*}{2\pi}} a_R(k) \sigma^+ \right] dk + \int_{-\infty}^0 \left[ \sqrt{\frac{V_L}{2\pi}} a_L^{\dagger}(k) \sigma^- + \sqrt{\frac{V_L^*}{2\pi}} a_L(k) \sigma^+ \right] dk,
\end{split}
\end{align}
where $\omega(k)$ is the waveguide dispersion relation, $a_R(k)$ and $a_L(k)$ are the bosonic waveguide mode operators for right- and left-moving photons with wave number $k$, respectively, $V_R$ ($V_L$) is the coupling rate between the emitter and the right-moving (left-moving) waveguide mode, and, as in our model, $\omega_{\rm e}$ is the transition frequency of the emitter and $\sigma^\pm$ are the raising and lowering operators. We have assumed that the coupling rates are independent of the photon wave number $k$.

\begin{figure}[t]
	\centering
	\includegraphics[width=0.6\linewidth]{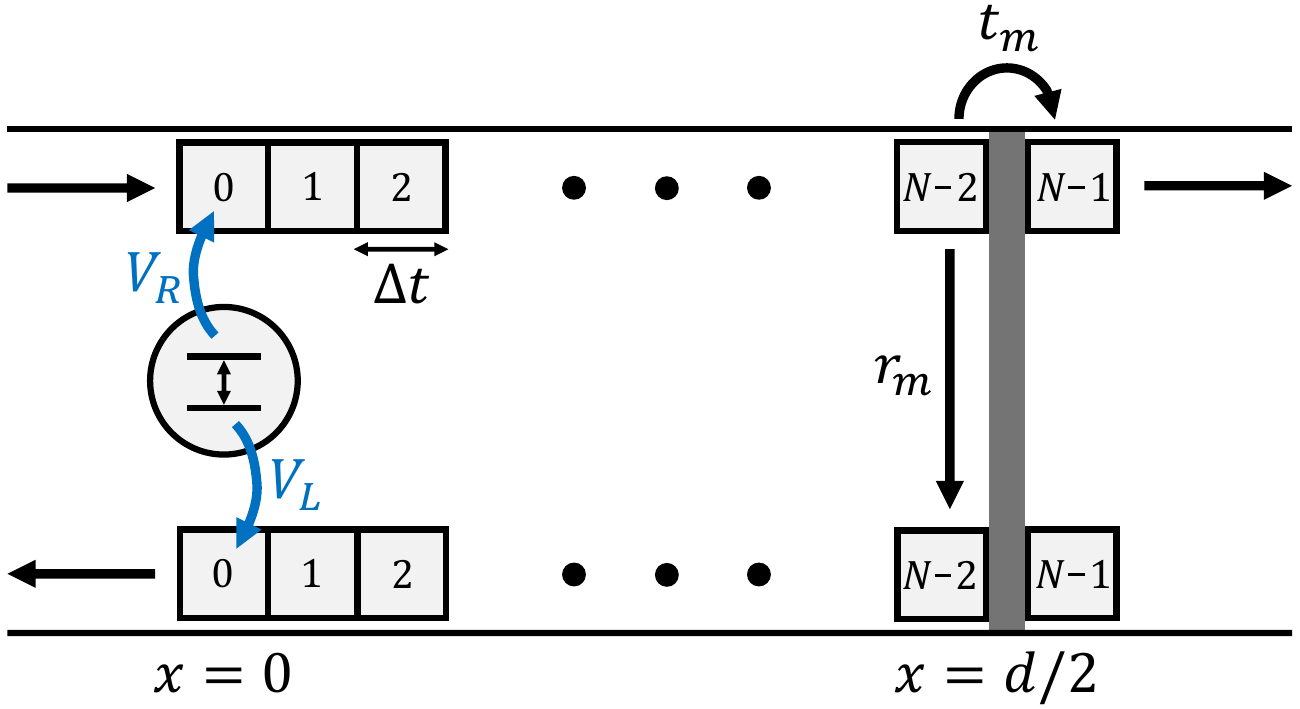}
	\caption{The emitter--mirror system within the space-discretized quantum trajectory model. The waveguide region of interest is divided into $N$ right-moving boxes and $N$ left-moving boxes, each having width $\Delta t$ in time. The boxes are labeled with the index ${n\in\{0,1,\ldots,N-1\}}$. The emitter couples to box 0 with coupling rate $V_R$ to the right and $V_L$ to the left. The boxes labeled with index ${N-1}$ are behind the mirror to account for transmission through the mirror.}
	\label{fig:trajectory_diagram}
\end{figure}

We discretize the $k$-space Hamiltonian in Eq.~(\ref{eq:H_k}) by replacing the integrals with summations over a finite number $N$ of $k$-space modes, and the continuum waveguide mode operators $a_{\mu}(k)$, $a_{\mu}^{\dagger}(k)$ with discrete mode operators $a_{\mu,k}$, $a_{\mu,k}^{\dagger}$ (${\mu\in\{L,R\}}$):
\begin{equation}
    H = \omega_{\rm e} \sigma^+ \sigma^- + \sum_{\mu=L,R} \sum_{k=0}^{N-1} \omega_k a_{\mu,k}^{\dagger} a_{\mu,k} + \sqrt{\frac{2\pi}{L_0}} \sum_{\mu=L,R} \sum_{k=0}^{N-1} \left( \sqrt{\frac{V_\mu}{2\pi}} a_{\mu,k}^{\dagger} \sigma^- + \sqrt{\frac{V_\mu^*}{2\pi}} a_{\mu,k} \sigma^+ \right),
\end{equation}
where ${\omega_k = 2\pi k/L_0}$ is the discretized (linear) dispersion relation, and ${L_0 = N\Delta t}$ is the length of the discretized region (we use the convention ${c = 1}$ for consistency with other authors, e.g., Ref.~\cite{Regidor2021}). This Hamiltonian is then divided into three parts:
\begin{equation}
    H = H_S + H_W + H_I,
\end{equation}
where
\begin{equation}\label{eq:H_S}
    H_S = \omega_{\rm e} \sigma^+ \sigma^-
\end{equation}
is the Hamiltonian of the two-level emitter,
\begin{equation}
    H_W = \sum_{\mu=L,R} \sum_{k=0}^{N-1} \omega_k a_{\mu,k}^{\dagger} a_{\mu,k}
\end{equation}
is the free waveguide Hamiltonian for the discretized region, and
\begin{equation}\label{eq:H_I}
    H_I = \sqrt{\frac{2\pi}{L_0}} \sum_{\mu=L,R} \sum_{k=0}^{N-1} \left( \sqrt{\frac{V_\mu}{2\pi}} a_{\mu,k}^{\dagger} \sigma^- + \sqrt{\frac{V_\mu^*}{2\pi}} a_{\mu,k} \sigma^+ \right)
\end{equation}
is the emitter--waveguide interaction.

To simulate the time evolution using this Hamiltonian, we need to express the Hamiltonian and any other observable of interest as a matrix by choosing an appropriate basis. The basis we use is a tensor product of the emitter basis $\{\ket{{\rm g}}, \ket{{\rm e}}\}$ and the photon number states of the $2N$ spatial boxes. Since we consider single-photon emission in our paper, each box is either unoccupied or occupied by one photon at any given time, and a maximum of one excitation can exist in the system. The total basis containing states with up to one excitation is:
\begin{align}
\begin{split}
\Bigl\{ &\ket{{\rm g}, 0_{R,0}, \ldots, 0_{R,N-1}, 0_{L,N-1}, \ldots, 0_{L,0}}, \\
   &\ket{{\rm e}, 0_{R,0}, \ldots, 0_{R,N-1}, 0_{L,N-1}, \ldots, 0_{L,0}}, \\[0.05in]
   &\ket{{\rm g}, 1_{R,0}, \ldots, 0_{R,N-1}, 0_{L,N-1}, \ldots, 0_{L,0}}, \\
   &\hspace{1.15in} \vdots\\
   &\ket{{\rm g}, 0_{R,0}, \ldots, 0_{R,N-1}, 0_{L,N-1}, \ldots, 1_{L,0}} \Bigr\},
\label{eq:basis}
\end{split}
\end{align}
which contains ${2N+2}$ elements. The first state is the vacuum state of the system, the second state corresponds to the emitter being excited, and the last $2N$ states correspond to a photon being present in one of the waveguide boxes. Here $\ket{0_{\mu,n}}$ ($\ket{1_{\mu,n}}$) corresponds to zero photons (one photon) being present in box $n$ moving in the $\mu$ direction.

In the trajectory simulations, we choose the initial state to be the same as in our model, where the emitter is excited and the waveguide is in the vacuum state [see Eq.~(\ref{eq:initial_state})]. In the above basis, this is written as:
\begin{equation}
    \ket{\psi(0)} = \ket{{\rm e}, 0_{R,0}, \ldots, 0_{R,N-1}, 0_{L,N-1}, \ldots, 0_{L,0}}.
\end{equation}
This is simply the second basis element, so the initial state can be represented by a vector of length ${2N+2}$ where the second element of the vector is one and the rest of the elements are zero. We evolve this state in time by applying the following algorithm at each time step:\\
\\
\indent
(1) Calculate the two-level emitter excitation probability
\begin{equation}
    P_{\rm e}(t) = \expval{\sigma^+ \sigma^-}{\psi(t)} = \left| \braket{{\rm{e}}| \psi(t)} \right|^2
\end{equation}
at the beginning of the time step, using the current state $\ket{\psi(t)}$ (any other observable of interest could also be calculated by taking the expectation value of the relevant operator).\\
\\
\indent
(2) Evolve the state $\ket{\psi(t)}$ by one time step $\Delta t$ under the sum of the two-level emitter Hamiltonian $H_S$ and the interaction Hamiltonian $H_I$:
\begin{equation}
    \ket{\psi(t+\Delta t)} = e^{-i(H_S + H_I)\Delta t} \ket{\psi(t)}.
\end{equation}

(3) Simulate a photon number measurement on the output boxes, and project the state according to the measurement outcome.\\
\\
\indent
(4) Shift the waveguide boxes along by one (i.e., evolve the state by one time step under the waveguide Hamiltonian $H_W$), taking into account transmission and reflection at the mirror.\\
\\
\indent
(5) Renormalize the state before the next iteration:
\begin{equation}
    \ket{\psi(t+\Delta t)} \rightarrow \frac{\ket{\psi(t+\Delta t)}}{\sqrt{\braket{\psi(t+\Delta t) | \psi(t+\Delta t)}}}.
\end{equation}

In order to implement steps (1) and (2), we need to find the matrix representations of $\sigma^+\sigma^-$, $H_S$, and $H_I$ in our basis, which allows us to express $e^{-i(H_S + H_I)\Delta t}$ as a matrix. Due to the size of the basis, we will have a ${(2N+2)\times(2N+2)}$ matrix in each case. From Eq.~(\ref{eq:H_S}), it follows that $H_S$ only has one nonzero matrix element in the second row and second column (the diagonal element where the emitter is excited), given by $\omega_{\rm e}$. The matrix for the observable $\sigma^+\sigma^-$ is the same as that for ${H_S = \omega_{\rm e} \sigma^+ \sigma^-}$, except with $\omega_{\rm e}$ replaced by one. To find the matrix representation of $H_I$, we need to use the following discrete Fourier transforms to replace the $k$-space waveguide operators in Eq.~(\ref{eq:H_I}) with position-space operators for the spatial boxes:
\begin{equation}
a_{R,k} = \frac{1}{\sqrt{N}} \sum_{n=0}^{N-1} A_{R,n} e^{-i \omega_k n \Delta t} \quad \text{and} \quad a_{L,k} = \frac{1}{\sqrt{N}} \sum_{n=0}^{N-1} A_{L,n} e^{i \omega_k n \Delta t},
\label{eq:inverse_fT}
\end{equation}
where $A_{\mu,n}$ is the annihilation operator for box $n$ moving in the $\mu$ direction. Substituting these Fourier transforms into Eq.~(\ref{eq:H_I}) leads to:
\begin{align}\label{eq:eq:H_I_position}
\begin{split}
    H_I =&\; \sqrt{\frac{2\pi}{L_0}} \sum_{k=0}^{N-1} \left[ \sqrt{\frac{V_R}{2\pi}} \left( \frac{1}{\sqrt{N}} \sum_{n=0}^{N-1} A_{R,n}^{\dagger} e^{i \omega_k n \Delta t} \right)\sigma^- + \sqrt{\frac{V_R^*}{2\pi}} \left( \frac{1}{\sqrt{N}} \sum_{n=0}^{N-1} A_{R,n} e^{-i \omega_k n \Delta t} \right) \sigma^+ \right]\\[0.05in]
    &+\sqrt{\frac{2\pi}{L_0}} \sum_{k=0}^{N-1} \left[ \sqrt{\frac{V_L}{2\pi}} \left( \frac{1}{\sqrt{N}} \sum_{n=0}^{N-1} A_{L,n}^{\dagger} e^{-i \omega_k n \Delta t} \right)\sigma^- + \sqrt{\frac{V_L^*}{2\pi}} \left( \frac{1}{\sqrt{N}} \sum_{n=0}^{N-1} A_{L,n} e^{i \omega_k n \Delta t} \right) \sigma^+ \right]\\[0.1in]
    =&\; \sqrt{\frac{2\pi}{N L_0}} \sum_{n=0}^{N-1} \left[ \sqrt{\frac{V_R}{2\pi}} A_{R,n}^{\dagger} \sigma^- \left( \sum_{k=0}^{N-1}  e^{i \omega_k n \Delta t} \right) + \sqrt{\frac{V_R^*}{2\pi}} A_{R,n} \sigma^+ \left( \sum_{k=0}^{N-1}  e^{-i \omega_k n \Delta t} \right)\right.\\[0.05in]
    &\left.\hspace{0.85in}+ \sqrt{\frac{V_L}{2\pi}} A_{L,n}^{\dagger} \sigma^- \left( \sum_{k=0}^{N-1} e^{-i \omega_k n \Delta t} \right) + \sqrt{\frac{V_L^*}{2\pi}} A_{L,n} \sigma^+ \left( \sum_{k=0}^{N-1} e^{i \omega_k n \Delta t} \right) \right].
\end{split}
\end{align}
Using ${\omega_k = 2\pi k/L_0}$ and ${L_0 = N\Delta t}$, we have
\begin{equation}
\sum_{k=0}^{N-1} e^{\pm i \omega_k n \Delta t} = \sum_{k=0}^{N-1} e^{\pm 2\pi i k n \Delta t / L_0} = \sum_{k=0}^{N-1} e^{\pm 2\pi i k n / N} = N \left[ \frac{1}{N} \sum_{k=0}^{N-1} e^{\pm 2\pi i k (n-0) / N} \right] = N \delta_{n0},
\label{eq:exp_sum}
\end{equation}
which follows from the identity
\begin{equation}
    \delta_{nm} = \frac{1}{N} \sum_{k=0}^{N-1} e^{\pm 2\pi i k (n-m) / N}.
\end{equation}
Using Eq.~(\ref{eq:exp_sum}) in Eq.~(\ref{eq:eq:H_I_position}) allows the sum over $n$ to be evaluated, leading to
\begin{align}
\begin{split}
    H_I =&\; \sqrt{\frac{N}{L_0}} \left( \sqrt{V_R} A_{R,0}^{\dagger} \sigma^- + \sqrt{V_R^*} A_{R,0} \sigma^+ + \sqrt{V_L} A_{L,0}^{\dagger} \sigma^- + \sqrt{V_L^*} A_{L,0} \sigma^+ \right)\\[0.05in]
    =& \sum_{\mu = L,R} \left( \sqrt{\frac{V_{\mu}}{\Delta t}} A_{\mu,0}^{\dagger} \sigma^- + \sqrt{\frac{V_{\mu}^*}{\Delta t}} A_{\mu,0} \sigma^+ \right).
\end{split}
\end{align}
The matrix elements of $H_I$ can now be found in the chosen basis using the way in which the emitter operators $\sigma^\pm$ and the position-space waveguide operators $A_{\mu,0}$ and $A_{\mu,0}^{\dagger}$ act on the basis states in Eq.~(\ref{eq:basis}) (for example, ${A_{\mu,0}^{\dagger}\ket{0_{\mu,0}} = \ket{1_{\mu,0}}}$ and ${A_{\mu,0}\ket{1_{\mu,0}} = \ket{0_{\mu,0}}}$).

In step (3), we simulate the photon number measurement by calculating the detection probabilities
\begin{equation}
    P_{\mu} = \braket{\psi_\mu|\psi_\mu}
\end{equation}
corresponding to the output boxes where the photon is detected (box ${N-1}$ for ${\mu=R}$ and box $0$ for ${\mu=L}$; see Fig.~\ref{fig:trajectory_diagram}). Here $\ket{\psi_R}$ ($\ket{\psi_L}$) is the projected state with all amplitudes except that of the basis state containing $\ket{1_{R,N-1}}$ ($\ket{1_{L,0}}$) set to zero. For $\ket{\psi_R}$ this is basis element ${N+2}$ and for $\ket{\psi_L}$ this is basis element ${2N+2}$ [see Eq.~(\ref{eq:basis})]. The total detection probability ${P_{\text{tot}} = P_R + P_L}$ is then compared against a uniformly-distributed random number ${\epsilon_1 \in (0,1)}$. If ${\epsilon_1 \leq P_{\text{tot}}}$ then a photon is detected, and the current state is projected as
\begin{equation}
    \ket{\psi(t+\Delta t)} \rightarrow \frac{\ket{\psi_{\mu}}}{\sqrt{\braket{\psi_{\mu}|\psi_{\mu}}}},
\end{equation}
where ${\mu=R}$ or ${\mu=L}$ is chosen by comparing the individual detection probabilities $P_R$ and $P_L$ against a second uniformly-distributed random number, ${\epsilon_2 \in (0,P_{\text{tot}})}$ (if ${\epsilon_2 \leq P_L}$ then project onto $\ket{\psi_L}$, otherwise project onto $\ket{\psi_R}$). On the other hand, if ${\epsilon_1 > P_{\text{tot}}}$ then no photon is detected in the output boxes, and the current state is instead projected according to
\begin{equation}
    \ket{\psi(t+\Delta t)} \rightarrow \frac{\ket{\psi_0}}{\sqrt{\braket{\psi_0|\psi_0}}},
\end{equation}
where $\ket{\psi_0}$ is the projected state with the amplitudes of the basis states containing $\ket{1_{R,N-1}}$ and $\ket{1_{L,0}}$ set to zero (elements ${N+2}$ and ${2N+2}$, respectively).

In step (4), we shift the waveguide boxes by moving the amplitudes in the vector $\ket{\psi(t+\Delta t)}$ accordingly, such that the amplitudes corresponding to a photon being in the right-moving boxes are moved away from box $0$ and towards box ${N-1}$, and the amplitudes corresponding to a photon being in the left-moving boxes are moved away from box ${N-1}$ and towards box $0$. To simulate the presence of the mirror, we take the amplitude of the basis state containing $\ket{1_{R,N-1}}$ to be the amplitude of the state containing $\ket{1_{R,N-2}}$ from the previous time step, multiplied by the transmission coefficient ${t_{\rm m} = \sqrt{1-|r_{\rm m}|^2}}$ (which we calculate from the reflection coefficient $r_{\rm m}$ and assume to be real, which is the case in the results we present). In addition, we take the amplitude of the state containing $\ket{1_{L,N-2}}$ to be the amplitude of the state containing $\ket{1_{R,N-2}}$ from the previous time step, multiplied by $r_{\rm m}$. We assume that the input box behind the mirror (left-moving box labeled ${N-1}$) is always empty, so nothing is transmitted or reflected from the other side of the mirror. This process of shifting boxes does not conserve the norm of the state ${\ket{\psi(t+\Delta t)}}$, so we renormalize the state in step (5) before the next time step begins.

Steps (1)-(5) are repeated until a chosen end time to obtain a single trajectory. The process of repeatedly measuring the system is stochastic, so we need to average over many trajectories to find the actual time evolution of the excitation probability $P_{\rm e}(t)$.


\subsection{Comparison of results}\label{app_subsec:comparison}

\begin{figure}
    \centering
    \includegraphics[width=0.5\linewidth]{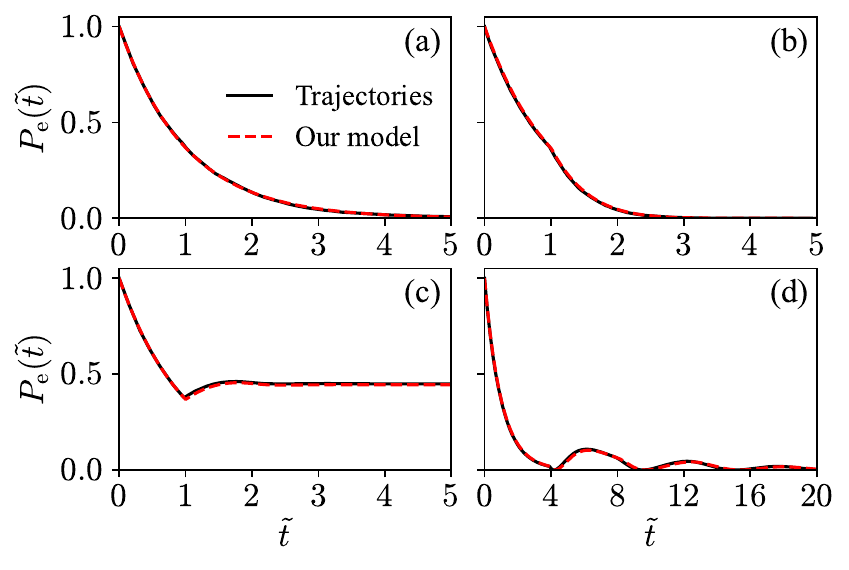}
    \caption{Comparison between the results of our analytical model (dashed red curves) and numerical quantum trajectory simulations (solid black curves), using some of the parameter sets from Fig.~\ref{fig:plots_non-Markovian}. The reflection coefficients are (a) ${r_{\rm m}=0}$, (b) ${r_{\rm m}=-0.5}$, and (c), (d) ${r_{\rm m}=-1}$. The round-trip phases are (b) ${\tilde{\omega}_{\rm e} \tilde{\tau} = \pi}$ (${\tilde{\tau} = 1}$), (c) ${\tilde{\omega}_{\rm e} \tilde{\tau} = 2\pi}$ (${\tilde{\tau} = 1}$), and (d) ${\tilde{\omega}_{\rm e} \tilde{\tau} = \pi}$ (${\tilde{\tau} = 4}$).}
    \label{fig:trajectories_comparison}
\end{figure}

The results of the quantum trajectory simulations are presented in Fig.~\ref{fig:trajectories_comparison} (solid black curves), with a comparison to the results of our analytical model (dashed red curves). For the comparison, we use some of the parameter sets that were used in Fig.~\ref{fig:plots_non-Markovian} to cover a range of different reflection coefficients, emitter--mirror separations, and round-trip phases. The trajectory results in Fig.~\ref{fig:trajectories_comparison} were obtained by averaging over $5000$ trajectories with ${N=25}$ boxes moving in each direction. The number of boxes determines the time step ${\Delta t = d/[2(N-1)c]}$ that needs to be used in the simulations, since ${d/2 = (N-1)c\Delta t}$ is the distance between the emitter and the mirror. The comparison with our model was made by using the same parameter normalization in both cases, i.e., all parameters are normalized to the total decay rate ${V_R + V_L \equiv \Gamma}$ (${= 2|g|^2/c}$ in our model). As Fig.~\ref{fig:trajectories_comparison} shows, in each case we find excellent agreement between our analytical results and the numerical simulations based on the more well-established quantum trajectory method. This demonstrates that our model can correctly describe the full non-Markovian dynamics of the emitter--mirror system.

We note that there are several advantages of our method compared to numerical approaches like quantum trajectories. First, our method provides exact results for the full dynamics of the quantum emitter and the photon wave packet, allowing the physics of the system to be studied analytically in different parameter regimes. This contrasts with numerical approaches where the dynamics converge to the exact result asymptotically (in the present case, when the number of trajectories ${N_T \rightarrow \infty}$). In addition, because our method is completely analytical, it provides an intuitive picture of the non-Markovian dynamics based on local interaction processes at the positions of the emitter and the mirror. Furthermore, the simple result in Eq.~(\ref{eq:Pe_markovian}) clearly shows how the mirror modifies the decay rate of an emitter in the Markovian limit, predicting interference-induced enhancement and suppression of emission that is analogous to the superradiance and subradiance effects observed with two coupled emitters.


\section{Calculation of $P_{\rm L}(x,t)$}\label{app:position}

In this appendix, we show how we obtain the probability density $P_{\rm L}(x,t)$ in Fig.~\ref{fig:wave_packets} from the state in Eq.~(\ref{eq:state_left}), which is the left-moving component of the field for ${x<0}$. We first separate the two terms in Eq.~(\ref{eq:state_left}) and write the (unnormalized) photon state $\ket{\psi_{\rm L}(t)} $ as 
\begin{align}
\begin{split}
     \ket{\psi_{\rm L}(t)} =& -ig \sum_{k=0}^{\infty} \int_0^t {\rm d}t_1 \;  e^{-i \Omega t_1} \frac{a^k}{k!} \left( t_1- kd/c \right)^{k} \; \Theta \left( t_1 - kd/c \right) \ket{{\rm g}, 1_{-1}(-c(t-t_1))} \\
     &-ig \sum_{k=0}^{\infty} \int_0^{t-d/c} {\rm d}t_1 \;  e^{-i \Omega t_1} \frac{a^k}{k!} \left( t_1- kd/c \right)^{k} \; \Theta \left( t_1 - kd/c \right) r_{\rm m} \ket{{\rm g}, 1_{-1}(d - c(t-t_1))},
\end{split}
\end{align}
where we used the step function ${\Theta( c(t-t_1) - d)}$ that was present in the reflected component to change the upper integration limit in the second line (the step function is one only if ${t_1 < t-d/c}$). Next, we use the substitution ${t_1 \rightarrow t_1 - d/c}$ in the second integral to obtain
\begin{align}\label{eq:state_t1}
\begin{split}
     \ket{\psi_{\rm L}(t)} =& -ig \sum_{k=0}^{\infty} \int_0^t {\rm d}t_1 \;  e^{-i \Omega t_1} \frac{a^k}{k!} \left( t_1- kd/c \right)^{k} \; \Theta \left( t_1 - kd/c \right) \ket{{\rm g}, 1_{-1}(-c(t-t_1))} \\
     &-ig \sum_{k=0}^{\infty} \int_{d/c}^{t} {\rm d}t_1 \;  e^{-i \Omega t_1} \frac{a^k}{k!} \left( t_1- (k+1)d/c \right)^{k} \; \Theta \left( t_1 - (k+1)d/c \right) r_{\rm m} e^{i \Omega d/c} \ket{{\rm g}, 1_{-1}(-c(t-t_1))}.
\end{split}
\end{align}
Due to the step functions in Eq.~(\ref{eq:state_t1}), we can change the lower integration limits to $kd/c$ and ${(k+1)d/c}$ in the first and second integrals, respectively:
\begin{align}\label{eq:state_t1_2}
\begin{split}
     \ket{\psi_{\rm L}(t)} =& -ig \sum_{k=0}^{\infty} \int_{kd/c}^t {\rm d}t_1 \;  e^{-i \Omega t_1} \frac{a^k}{k!} \left( t_1- kd/c \right)^{k} \ket{{\rm g}, 1_{-1}(-c(t-t_1))} \\
     &-ig \sum_{k=0}^{\infty} \int_{(k+1)d/c}^{t} {\rm d}t_1 \;  e^{-i \Omega t_1} \frac{a^k}{k!} \left( t_1- (k+1)d/c \right)^{k} \; r_{\rm m} e^{i \Omega d/c} \ket{{\rm g}, 1_{-1}(-c(t-t_1))}.
\end{split}
\end{align}
We now use the substitution ${x = -c(t-t_1)}$ in both integrals in Eq.~(\ref{eq:state_t1_2}), which leads us to
\begin{align}\label{eq:state_x}
\begin{split}
     \ket{\psi_{\rm L}(t)} =& -\frac{ig}{c} \sum_{k=0}^{\infty} \int_{kd-ct}^0 {\rm d}x \;  e^{-i \Omega (x/c+t)} \frac{a^k}{k!} \left( x/c + t - kd/c \right)^{k} \ket{{\rm g}, 1_{-1}(x)} \\
     &-\frac{ig}{c} \sum_{k=0}^{\infty} \int_{(k+1)d-ct}^0 {\rm d}x \;  e^{-i \Omega (x/c+t)} \frac{a^k}{k!} \left( x/c + t - (k+1)d/c \right)^{k} \;  r_{\rm m} e^{i \Omega d/c} \ket{{\rm g}, 1_{-1}(x)}.
\end{split}
\end{align}
This represents a wave packet that extends from ${x=-ct}$ to ${x=0}$, where the emitter is located.

The lower integration limits in Eq.~(\ref{eq:state_x}) mean that we need to partition the spatial region into intervals of length $d$. In the first interval, ${-ct \leq x < -ct+d}$, only the ${k=0}$ term in the first line contributes, and is given by
\begin{equation}\label{eq:state_x_1}
    -\frac{ig}{c} \int_{-ct}^{-ct+d} {\rm d}x \;  e^{-i \Omega (x/c+t)} \ket{{\rm g}, 1_{-1}(x)}.
\end{equation}
This term has an exponential profile, as for a photon emitted in an infinite waveguide with no mirror. This is expected because the field component that is reflected from the mirror has not had time to return from the mirror and interfere with the field component emitted directly to the left. In the second interval, ${-ct + d\leq x < -ct+2d}$, the ${k=0}$ and ${k=1}$ terms from the first line in Eq.~(\ref{eq:state_x}) contribute, and the ${k=0}$ term from the second line contributes:
\begin{equation}
    -\frac{ig}{c} \int_{-ct+d}^{-ct+2d} {\rm d}x \;  e^{-i \Omega (x/c+t)} \left[ 1 + a(x/c+t-d/c) + r_{\rm m} e^{i \Omega d/c} \right] \ket{{\rm g}, 1_{-1}(x)}.
\end{equation}
More generally, in the interval ${-ct+nd \leq x < -ct+(n+1)d}$ with ${n\geq1}$ we can write the state as
\begin{equation}\label{eq:state_x_n}
    -\frac{ig}{c} \int_{-ct+nd}^{-ct+(n+1)d} {\rm d}x \;  e^{-i \Omega (x/c+t)} \phi_{\rm L}(x,t) \ket{{\rm g}, 1_{-1}(x)},
\end{equation}
where
\begin{equation}\label{eq:amplitude_x}
    \phi_{\rm L}(x,t) = \sum_{k=0}^n \frac{a^k}{k!}(x/c+t-kd/c)^k + r_{\rm m} e^{i \Omega d/c} \sum_{k=0}^{n-1} \frac{a^k}{k!}(x/c+t-(k+1)d/c)^k.
\end{equation}
The probability density $P_{\rm L}(x,t)$ in the interval ${-ct+nd \leq x < -ct+(n+1)d}$ is then given by the modulus-squared of the amplitude
\begin{equation}
    \Phi_{\rm L}(x,t) = -\frac{ig}{c} e^{-i \Omega (x/c+t)} \phi_{\rm L}(x,t)
\end{equation}
of the $\ket{{\rm g},1_{-1}(x)}$ state in Eq.~(\ref{eq:state_x_n}):
\begin{equation}
    P_{\rm L}(x,t) = \left| \Phi_{\rm L}(x,t) \right|^2 = \frac{|g|^2}{c^2} e^{-\Gamma (x/c+t)} \left|\phi_{\rm L}(x,t)\right|^2,
\end{equation}
as in Eq.~(\ref{eq:P_L}). Note that, in the first interval where ${-ct \leq x < -ct+d}$, we have ${P_{\rm L}(x,t) = \frac{|g|^2}{c^2} e^{-\Gamma (x/c+t)}}$, which follows from Eq.~(\ref{eq:state_x_1}). The final interval is ${-d \leq x < 0}$, where ${n=ct/d-1}$ in Eqs.~(\ref{eq:state_x_n}) and (\ref{eq:amplitude_x}).

\end{widetext}


%

\end{document}